\documentclass[reprint,superscriptaddress, amsmath,amssymb, aps, floatfix, showkeys]{revtex4-1}

\usepackage{cmap}
\usepackage{url}
\usepackage{newtxtext}
\usepackage[slantedGreek]{newtxmath}
\newcommand{\mathbold}[1]{\ensuremath{\textit{\textbf{#1}}}}

\usepackage{helvet}

\usepackage{mleftright}
\usepackage{graphicx}%
\usepackage{bm}%
\usepackage{graphicx}%
\usepackage{xcolor}
\usepackage[version=4]{mhchem}
\usepackage{comment}
\usepackage{siunitx}
\DeclareSIUnit{\calorie}{cal}

\usepackage[stretch=15,shrink=15,step=1]{microtype}

\usepackage[pdftex, bookmarks, pdffitwindow=false, pdfstartview=FitH, pdfdisplaydoctitle, colorlinks, plainpages=false, pdftitle={Predicting the phase diagram of titanium dioxide with random search and pattern recognition},pdfauthor={Aleks Reinhardt, Chris Pickard, Bingqing Cheng}, pdfpagelabels, hypertexnames, citecolor={cyan},linkcolor={cyan}, urlcolor={violet}, pdflang={en}, hyperfootnotes=false, breaklinks]{hyperref}

\newcommand{\silabel}{Supplementary Information}
\newcommand{\methodsLink}{\hyperref[sec:methods]{Methods}}

\newcommand{\rev}[1]{{\color{black} #1}}

\newcommand{\abs}[1]{\ensuremath{\lvert {#1}\rvert}}

\newcommand{\kb}{k_{\rm B}} %

\newcommand{\titaniaForm}{TiO$_2$}  %
\newcommand{\figref}[1]{Fig.~\hyperref[#1]{\ref{#1}}}
\newcommand{\figrefsub}[2]{Fig.~\hyperref[#1]{\ref{#1}(#2)}}

\makeatletter
\def\@bibdataout@aps{
 \immediate\write\@bibdataout{
 @CONTROL{
   apsrev41Control, author="48",editor="1",pages="1",title="0",year="1"
 }}
 \if@filesw
  \immediate\write\@auxout{\string\citation{apsrev41Control}}
 \fi
}
\makeatother

\usepackage[nodayofweek]{datetime}
\newdateformat{myDate}{\THEDAY\ \monthname[\THEMONTH] \THEYEAR} %

\begin{document}

\title{Predicting the phase diagram of titanium dioxide with random search and pattern recognition}
\author{Aleks Reinhardt}
\affiliation{Department of Chemistry, University of Cambridge, Lensfield Road, Cambridge, CB2 1EW, United Kingdom}

\author{Chris J. Pickard}
\affiliation{Department of Materials Science \& Metallurgy, University of Cambridge, 27 Charles Babbage Road,
Cambridge, CB3 0FS, United Kingdom}
\affiliation{Advanced Institute for Materials Research, Tohoku University, Sendai, Japan}

\author{Bingqing Cheng}
\email{bc509@cam.ac.uk}
\affiliation{Department of Chemistry, University of Cambridge, Lensfield Road, Cambridge, CB2 1EW, United Kingdom}
\affiliation{TCM Group, Cavendish Laboratory, University of Cambridge, J.~J.~Thomson Avenue, Cambridge, CB3 0HE, United Kingdom}
\affiliation{Trinity College, Trinity Street, Cambridge, CB2 1TQ, United Kingdom}%

\date{\myDate\today}%

\begin{abstract}

Predicting phase stabilities of crystal polymorphs is central to computational materials science and chemistry. \rev{Such predictions are challenging
because they first require searching for potential energy minima
and then performing arduous free-energy calculations to account for entropic effects at finite temperatures.
Here, we develop a framework that facilitates such predictions
by exploiting all the information obtained from random searches of crystal structures.
This framework combines automated clustering, classification and visualisation of crystal structures with machine-learning estimation of their enthalpy and entropy.
We demonstrate the framework on the technologically important system of \titaniaForm{}, which has many polymorphs, without relying on prior knowledge of known phases. We find a number of new phases and predict the phase diagram and metastabilities of crystal polymorphs at 1600 K, benchmarking the results against full free-energy calculations.
}
\end{abstract}

\keywords{
Phase diagram; Entropy; Machine learning;
Structure prediction; Titanium dioxide}

\maketitle

\section{\label{sec:intro}Introduction}

Predicting the properties of solid materials requires an understanding of how atoms are arranged,
which in turn necessitates the determination of the relative stabilities of possible crystal polymorphic phases under various conditions.
Advances in crystal structure prediction~\cite{woodley2008crystal},
including random structure search (RSS)~\cite{pickard2011ab},
particle-swarm optimisation~\cite{wang2010crystal},
Monte Carlo simulations with variable box shapes~\cite{filion2009efficient}
and basin hopping~\cite{wales1999global},
allow us to find minima on the potential-energy surface (PES) and thus to identify potentially competitive polymorphs.
However, in these approaches
only the enthalpy $H$ at \SI{0}{\kelvin} is typically computed and used to determine phase stability,
even though the difference in entropy $S$ between polymorphs can play a significant role~\cite{cheng2018computing,Fu2013}.
Relative Gibbs energies ($G=H-TS$), which in actuality dictate thermodynamic (meta)stability, are seldom computed, 
despite their importance for phases that can be synthesised at one thermodynamic condition but remain metastable under other conditions.

Although Gibbs energies can be computed using free-energy methods such as thermodynamic integration (TI)~\cite{cheng2018computing,Vega2008, Reinhardt2019},
such calculations are non-trivial and usually require long simulations,
making them both expensive and tedious if one wants to use the PES computed from first-principles methods such as density-functional theory (DFT)~\cite{monserrat2018structure}.
Moreover,
crystal-structure prediction schemes typically result in a multitude of structures.
It can be challenging to rationalise how the structures are related to one another
and which ones are promising candidates under given thermodynamic conditions. 

In this study,
we propose an approximate method that combines RSS and machine learning (ML) to estimate free energies of solid phases and hence the thermodynamic phase behaviour of a system of interest with relatively little computational effort and without relying on prior knowledge of the known phases.
We have selected titanium dioxide at ambient and high pressures to benchmark the method
because it is a widely used metal oxide with a large number of polymorphs (see the \silabel{}),
many of which may have interesting optical~\cite{Dekura2011}, mechanical~\cite{Dubrovinsky2001} and electrochemical~\cite{Mukai2017} properties.
Furthermore, high-pressure \titaniaForm{} phases can also serve as analogues of the structures adopted by many other important \ce{AX2} systems that are of particular interest in geology, such as high-pressure silicas~\cite{StaunOlsen1999}.

\section{Methods}\label{sec:methods}

\paragraph{Calculations using empirical potential}
For RSS and accurate free energy calculations,
we focus on the simple MA empirical pair potential for \titaniaForm{}~\cite{Matsui1991} because of its low computational cost~\cite{Reinhardt2019}.

We performed RSS at \SI{0}{\giga\pascal}, \SI{20}{\giga\pascal}, \SI{40}{\giga\pascal} and \SI{60}{\giga\pascal} using the AIRSS~\cite{Pickard2006, Pickard2011}
package interfaced with \textsc{Lammps}~\cite{Plimpton1995}.
For each RSS run, we first chose a reasonable cell shape at random, and added 2, 3, 4, 6, 8, 9 or 12 formula units of \titaniaForm{} into the simulation cell at random positions while keeping the initial density of the cell close to the typical density range of this system.
We set a lower bound on the interatomic
distance for each pair of atomic species, but otherwise imposed no additional constraints or symmetries on the initial structures.
We then relaxed each structure using a second-order conjugate gradient algorithm until the forces on atoms and the difference between the target and actual pressures both became negligible.

To benchmark how well our approximate framework can predict the phase behaviour of \titaniaForm{} at ambient and high pressures as described by the MA potential,
we also performed free-energy calculations for both the known phases and 
the new phases obtained from RSS. 
Specifically, we computed the chemical potentials of the polymorphs using Frenkel--Ladd integration~\cite{Frenkel1984, Vega2008} and subsequent TI along isobars and isotherms, as detailed in Ref.~\citenum{Reinhardt2019}.

All the necessary input files for performing the above-mentioned calculations using AIRSS and \textsc{Lammps} are provided in the supporting data. %

\paragraph{DFT calculations}
For the metastable polymorphs identified with the MA potential,
we performed geometrical optimisation and computed the enthalpies at the DFT level over a pressure range of 
\SIrange{0}{70}{\giga\pascal} in steps of \SI{10}{\giga\pascal}.
We separately employed three common functionals, LDA~\cite{jones1989density}, PBE~\cite{perdew1996generalized} and PBEsol~\cite{perdew2008restoring}, using the CASTEP \textit{ab initio} simulation package~\cite{clark2005first}.
Full details of the DFT set-ups and configurations can be found in the input files supplied in the \silabel{}.
The key value of the current study lies in the data analysis, 
which we describe in the following section.
A Python notebook implementing the machine learning and data processing steps, input and data files are available on the public repository \url{https://github.com/BingqingCheng/TiO2_random_search_pattern_recognition}.

\section{Results and analysis}

\subsection{Characterisation of structures found by RSS}

At each pressure, RSS using the MA potential produced thousands of distinct \titaniaForm{} structures with different atomic co-ordinates, cell shapes and numbers of formula units in the cell.
Even though knowledge of the space groups, molar volumes and energies of the structures provides hints on how to classify them,
it is still a formidable task to sort through them manually.
In recent years,
ML-inspired approaches have been used for 
the classification and visualisation of atomic structures~\cite{ceriotti2011simplifying,engel2018mapping,de+16pccp,engel2018mapping,anelli2018generalized,Mavracic2018},
but they have not been systematically exploited in the context of recognising the metastable polymorphs from RSS, especially for the purpose of determining the phase behaviour of crystalline systems.
We have therefore developed and employed a ML-based method to compare and cluster the structures automatically.

\paragraph{Similarity measurements}
This pattern recognition task is built around the construction of a kernel matrix
$\{ K (A,B)\}$ that measures the similarity between each pair of structures $A$ and $B$ in the data set.
The kernel matrix should be positive-definite and normalised.
If it is not already normalised by construction, then
$K(A,B)$ should be divided by $\sqrt{K(A,A)K(B,B)}$.
The kernel function $K(A,B)$ can be formulated as an inner product of the features
of $A$ and $B$,
\begin{equation}
    K(A,B) = \Phi(A)^{\mathsf{T}} \Phi(B)      %
    = \sum_{i=1}^{M} \phi_i(A) \phi_i(B),
    \label{eqn:kmatrix}
\end{equation}
where $\Phi=\{\phi_i\}$ denotes the set of global fingerprints associated with the whole structure, rather than with individual atomic environments $\mathcal{X}_n^A$ that are centred on each atom $n$ in the structure $A$.
A straightforward way of obtaining these global fingerprints from
the local ones
is to take the average~\cite{de+16pccp},
\begin{equation}
    \Phi (A) = \dfrac{1}{N_A}
    \sum_{n=1}^{N_A} \Psi (\mathcal{X}_n^A),
\label{eq:fp-global}
\end{equation}
where $N_A$ denotes the total number of atoms in the system $A$ if it only contains one element.
If the system contains multiple atomic species and
if the local fingerprints are specific to atomic species,
the average in Eq.~\eqref{eq:fp-global} should be taken
over atoms of the same atomic species $\alpha$
and the resulting vectors associated with each species are then concatenated.
Apart from the average kernel used here, one may also use the MATCH or RE-MATCH kernel of Ref.~\cite{de+16pccp},
although we did not notice a significant improvement during the subsequent analysis.
We employ the Smooth Overlap of Atomic Positions (SOAP) framework that was introduced in Ref.~\cite{bart+13prb} to construct the local fingerprints $\Psi (\mathcal{X})$.
SOAP has been used together with Gaussian Process Regression in numerous applications, including metals, semiconductors, molecular crystals and small organic molecules~\cite{deri-csan17prb,de+16pccp,szla+14prb,bart+17sa,musi+18cs},
as well for structural identifications of bulk materials, including ice~\cite{engel2018mapping,anelli2018generalized} and \titaniaForm{}~\cite{Mavracic2018}.
The SOAP representation is specific to atom species.
For atoms of species $\alpha$ inside a local environment $\mathcal{X}$, 
it uses a smooth atomic density function
\begin{equation}
    \rho_{\mathcal{X}}^{\alpha}(\mathbold{r}) = \sum_{i \in \mathcal{X}_\alpha} \exp\mleft(-\frac{\left[\mathbold{r}-\mathbold{r}_i\right]^2}{2\sigma^2}\mright)
\end{equation}
by summing over Gaussians centred on each atom $i$ of species $\alpha$ that has a displacement $\mathbold{r}_i$ within a given cutoff $r_\text{c}$ of the central atom of the environment $\mathcal{X}$.
The density $\rho_{\mathcal{X}}^{\alpha}(\mathbold{r})$ is invariant to translations and permutations of identical atoms, but not to rotations.
The SOAP representation addresses this by
first expanding with a set of orthonormal basis functions on radial direction $g(\abs{r})$ and
spherical harmonics of angular directions $\hat{r}$ as
\begin{equation}
    \rho_{\mathcal{X}}^{\alpha}(\mathbold{r}) =
    \sum_{nlm} c_{nlm}^{\alpha} g_n(\abs{r}) Y_{lm} (\hat{r}),
\end{equation}
and then taking the power spectra that characterise the rotational-invariant arrangement of atoms of species $\alpha$ inside the local environment $\mathcal{X}$~\cite{bart+13prb,de+16pccp}
\begin{equation}
    k_{nn'l}^{\alpha} (\mathcal{X}) =
    \uppi \sqrt{\dfrac{8}{2l+1}}
    \sum_m (c_{nlm}^{\alpha})^* c_{n'lm}^{\alpha}.
    \label{eqn:partial-k}
\end{equation}
The vector $\{  k_{nn'l}^{\alpha}\}$ constructed in this way up to certain cutoffs $l_\text{max}$ and $n_\text{max}$
can then be used as the local fingerprint $\Psi (\mathcal{X})$ in Eq.~\eqref{eq:fp-global},
which in turn leads to the kernel matrix. %
In practice, we use the recently implemented DScribe Python package for constructing descriptors and kernels~\cite{dscribe}.

\begin{figure*}[hbt]
\centering
\includegraphics[trim=120 85 110 85, width=1.0\textwidth]{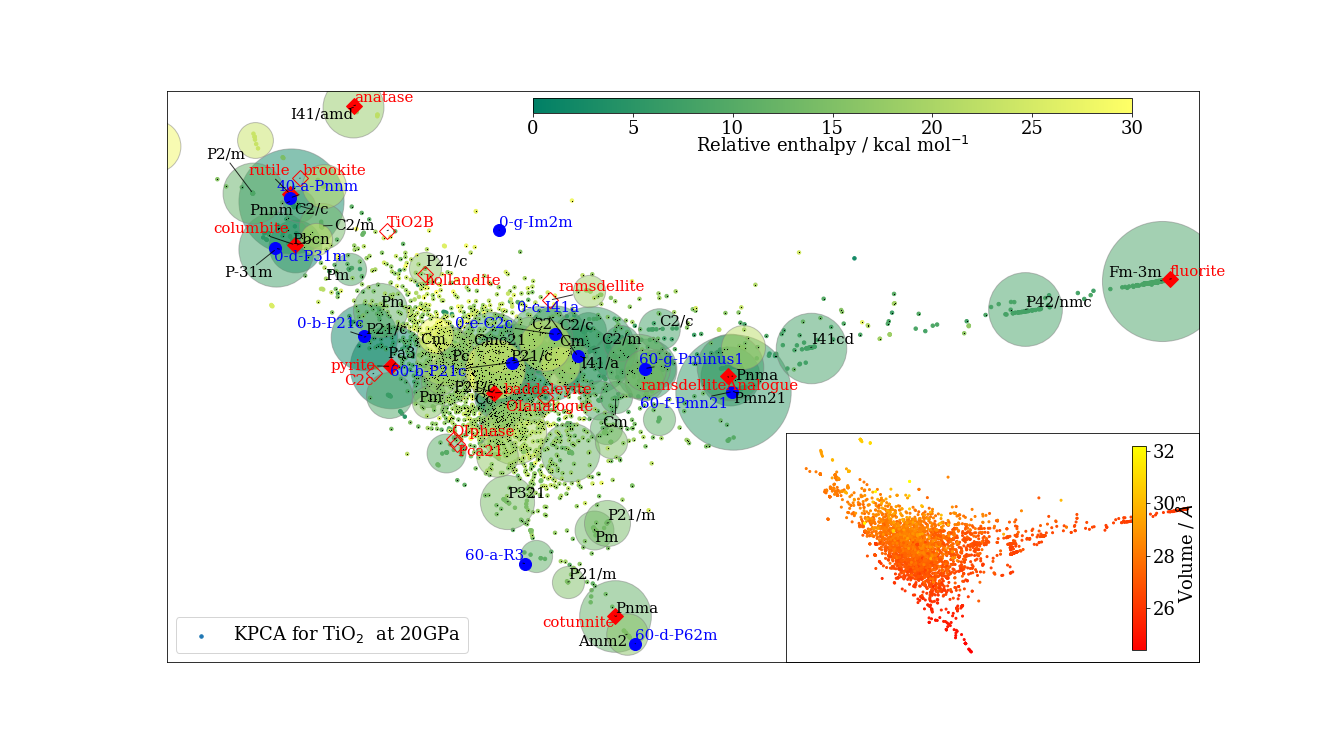}
\caption{
Output after KPCA and clustering analysis.
The structures obtained from RSS at \SI{20}{\giga\pascal} are projected onto the first two principal vectors of the kernel matrix $\{ K (A,B)\}$.
The known and new phases of \titaniaForm{} are indicated on the KPCA plot using solid markers if found during RSS, and hollow markers otherwise.
The colours of the points in the main graph and the inset vary according
to their enthalpy and volume per \titaniaForm{} formula unit, respectively.
The area of each marker is proportional to the logarithm of the number of the structures in that cluster.}
\label{fig:kpca}
\end{figure*}

\paragraph{Low-dimensional maps}
The kernel matrix
$\{ K (A,B)\}$ provides distance measurements between structures in a high-dimensional space.
To visualise such distances, we project them onto a two-dimensional map using
kernel principal component analysis (KPCA)~\cite{scholkopf1998nonlinear},
which amounts to projecting $\{ K (A,B)\}$ onto its two eigenvectors with the largest eigenvalues.
Fig.~\ref{fig:kpca} shows such 2D maps for the structures found during the RSS at \SI{20}{\giga\pascal}.

\paragraph{Clustering}
For the set of structures generated by RSS at each pressure,
we clustered them based on the similarity measurements
$\{ K (A,B)\}$
in order to find which structures belong to the same crystallographic family
using the Density-Based Spatial Clustering of Applications with Noise (DBSCAN) algorithm~\cite{ester1996density}.
In this approach, the density of data points around a given data point is first estimated and then points with a density above a certain threshold are identified as clusters.
Input data points
that are not ascribed to regions of high data density are classified as noise.
The most important DBSCAN parameter is the maximum distance between two samples for them to be considered  neighbours.
In our case, there is a very clear separation between similar and dissimilar structures obtained from RSS due to the absence of thermal noise,
so the clustering outcome is insensitive to this DBSCAN parameter.

As an example, in Fig.~\ref{fig:kpca}
we show such clustering results.
Even though the information on space groups, molar volume and enthalpy is not explicitly included in the construction of the kernel matrix
$\{ K (A,B)\}$,
this agnostic pattern recognition approach is able to
group together structures with similar properties extremely well.

\paragraph{Identification of known and new phases}

In the KPCA map of Fig.~\ref{fig:kpca},
we also project the location of previously found crystal structures of \titaniaForm{},
and, in an \textit{ad hoc} manner, the new structures found in the present study.
If a cluster that contains several structures found in RSS
also includes a previously known phase,
it is given the label of this phase,
otherwise it is regarded as a new phase.
In this way, we found 15 phases which have not to our knowledge previously been considered in studies of titanium dioxide in this pressure range.
In the remainder of this manuscript we label them with a standard format \texttt{P-a-SYM}, where \texttt{P} gives the pressure in gigapascals at which the structure was found in RSS, \texttt{a} is a lowercase letter purely used for labelling, and \texttt{SYM} gives the space group of the structure.
These new structures are illustrated and described in the \silabel{} (Fig.~S1).
 One of the new phases (60-d-P$\overline{6}$2\textit{m}) is in fact the Fe$_2$P phase of \titaniaForm{} that was previously considered at higher pressures than we have focused on here~\cite{Dekura2011,lyle2015prediction}.  

\paragraph{DFT calculations}
Although the MA potential satisfactorily predicts many properties of \titaniaForm{}~\cite{Collins1996b,Swamy2001},
it does not give a perfect description of interactions, so the fact that a new phase is found using the MA potential does not necessarily mean that the same holds for \titaniaForm{} in experiment.
To check whether the new structures may be of experimental relevance, we computed the lattice energies and enthalpies of all the phases at the DFT level over a pressure range of \SIrange{0}{70}{\giga\pascal} (see \methodsLink{}).
We used three functionals, LDA, PBE and PBEsol, and the results are shown in the \silabel{} (Fig.~S3).
Although a recent study~\cite{trail2017quantum} shows good agreement in the ranking of static-lattice energies of phases at low pressure between diffusion quantum Monte Carlo and a number of DFT functionals,
the DFT calculations for \titaniaForm{} should be approached with a degree of caution particularly at low pressure~\cite{lyle2015prediction}, since for example rutile is not predicted to be the stable phase at sufficiently low pressures for any of them, the results for the known phases are consistent with previous work~\cite{Ma2009, Fu2013, Mei2014, Zhu2014}.
At high pressures ($\sim$\SI{60}{\giga\pascal}),
the three functionals gave consistent results, suggesting that the results are robust under such conditions.
Although none of the newly reported phases have the lowest enthalpy, several phases are very competitive.
In particular, 60-d-P$\overline{6}$2\textit{m} and 60-e-P2$_1$/\textit{m} have  enthalpies very close to the ground-state phase of cotunnite at high pressures ($P\ge \SI{40}{\giga\pascal}$),
and the 60-a-R$\overline{3}$ phase, which, as we discuss below, is particularly favoured for the MA potential, 
has an enthalpy only a few \si{\kilo\calorie\per\mole} per formula unit larger than cotunnite.
The difference in relative enthalpy between different levels of theory highlights the importance of capturing metastable phases,
as in general the enthalpies predicted by empirical potentials or density-functional approximations may not be accurate enough to determine the true ground-state enthalpy.

\subsection{Approximation of entropy using frequency of appearance\label{sec:sb}}

\begin{figure}
\includegraphics[trim=3mm 15mm 18mm 26mm,clip,width=1.0\columnwidth]{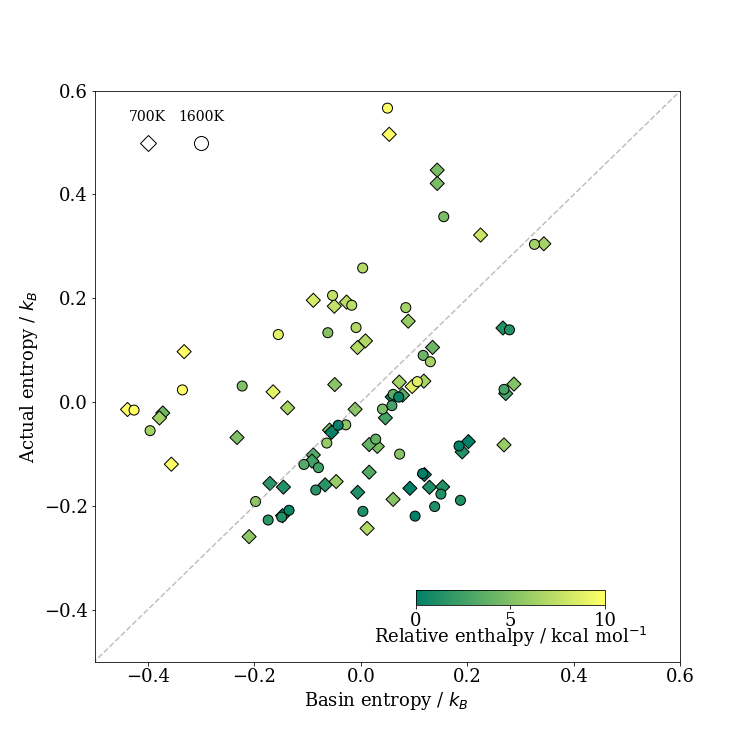}
\caption{
A comparison between the basin entropy $s_\text{b} = \kb \ln(v)/ n$ and the thermodynamic entropy $S = (H-G)/T$ per formula unit of \titaniaForm{}. %
Diamonds show results at \SI{700}{\kelvin} and circles show results at \SI{1600}{\kelvin}. 
Results obtained at all four pressures considered are all included,
and the label of each point can be found in the \silabel{} (Fig.~S4).
Each data point is coloured according to the relative enthalpy of a structure compared to the ground state at the same pressure.
As only the relative entropy under a given thermodynamic condition between different phases matters,
we centred the data about the origin.
}
\label{fig:entropy-basin}
\end{figure}

In this section, we investigate whether it is possible to approximate the entropies of different phases solely from the information one can obtain from RSS.
The RSS scheme, which involves performing energy minimisation of random structures, is rather similar to the direct enumeration method~\cite{xu2005random} that can be used to find the basin of attraction of granular or glassy systems~\cite{xu2005random,asenjo2014numerical,martiniani2017numerical},
although RSS is somewhat more complicated as variable cells with different numbers of atoms are used (see \methodsLink{}).
\rev{The entropy associated with the (dimensionless) volume $v$ of a basin of attraction, $S=k_\text{B}\ln v$, has been of considerable recent interest~\cite{Martiniani2016, *Daza2016}.} It would therefore be intriguing to explore whether in this case, we can find an estimate for the basin volume and whether the corresponding basin entropy is related to the thermodynamic entropy that enters the Gibbs energy.

To estimate the entropy of the basin of the crystal structures found in RSS, we assume that
(i) $S_\text{b}$ of a crystal structure is extensive with respect to its system size, $S_\text{b} = n s_\text{b}$, where $n$ is the number of \titaniaForm{} formula units; and
(ii) when performing energy minimisation with the same number of \titaniaForm{} formula units,
$n s_\text{b}$ is proportional to the logarithm of the frequency of finding a certain crystal structure.  
With these strong assumptions, we are able to infer the relative basin entropy for each crystal structure considered.
The value of $s_\text{b}$ obtained in this way depends on the frequency of occurrence of each structure for a given choice of $n$.
However, for structures whose unit cells do not have compatible numbers of formula units, we can nevertheless infer the difference in $s_\text{b}$ between two such polymorphs provided that they are found in searches together with other structures that have a well-defined $s_\text{b}$ for both values of $n$.

To test our assumptions, and to
compare the basin entropy $s_\text{b}$ of the crystal structures and the thermodynamic entropy $S$, we plot them against each other in Fig.~\ref{fig:entropy-basin}.
Given that the two sets of entropies were estimated using completely different approaches, it is remarkable how similar their spread is.
The Pearson correlation coefficient (PCC) is estimated to be $r\approx 0.3$,
and the root mean squared error (RMSE) is $0.2 \kb$.
\rev{The value of the PCC may at first glance seem low, but} note that the entropy difference between different phases of this system is itself relatively small,
so even a rather small RMSE can significantly reduce the PCC.
\rev{Moreover, } we can notice that low enthalpy structures tend to have $s_\text{b}$ larger than $S$: low enthalpy phases are found more frequently in RSS than one might expect from their thermodynamic entropies.
This may be because the way the initial structures are prepared and the subsequent enthalpy minimisation is performed during RSS can introduce a bias towards locating deeper minima,
which is advantageous if one aims to find stable phases over high energy ones.
\rev{The trend of finding more frequently the low enthalpy structures has been observed in a previous study that also involves sampling polymorphs of ionic solids~\cite{Stevanovi2016}.}
To achieve a better estimate of $S$, one can thus use a linear combination such as $s_\text{b}-aH$, where %
$a$ is a parameter to the fit $aH + \text{constant}= s_\text{b} - S$ over all data points.
We show a comparison between this adjusted entropy $s_\text{b}-aH$ and $S$ in Fig.~S4 of the \silabel{},
which shows a much stronger correlation ($r\approx 0.7$; \rev{see Section~S4}).
Of course, determining the value of $a$ requires prior knowledge of $S$, so such a correction scheme is less useful in practice.

Nevertheless, 
the results here demonstrate that the frequency of finding a structure in RSS does encode some information about its entropy.
Furthermore, the correlation between $s_\text{b}-S$ and $H$ means that
the difference in $s_\text{b}$ is a better estimator of the actual entropy difference between structures with similar enthalpies than for pairs with large enthalpy differences.
Since we are typically interested in phases in the vicinity of the most favourable structure, this may be sufficient. %

In order to demonstrate the level of approximation one can achieve in estimating the free energy using only RSS results, whenever we discuss ML-based approximations of $G$, we use $s_\text{b}$ to estimate the true entropy.
To obtain a more accurate estimate of $S$,
one can use the harmonic approximation~\cite{cheng2018computing},
or if resources allow,
the TI method~\cite{cheng2018computing,Vega2008, Reinhardt2019}.

\subsection{Machine-learning prediction of enthalpy and entropy}\label{subsect:ML-energetics}

As the number of local minima in the PES grows exponentially with increasing system size, it is difficult to ensure that all the important structures have been found during RSS.
Indeed, not all the previously characterised or new structures of \titaniaForm{} have been found at each pressure considered.
There may also be situations where a certain crystal structure is known for an analogous system, but may not have been considered for the system of interest.
In all such cases, we may wish to predict the enthalpies as well as the free energies of such structures.
Of course, for a certain lattice arrangement, one can choose to perform energy minimisations and free-energy calculations, but it would be convenient to have a surrogate model that can quickly assess its enthalpy and basin entropy.

We developed such an approximate model using the kernel ridge regression (KRR) approach.
In the following, we only discuss estimating $H$; the estimation of $S_\text{b}$ follows the same workflow.
For a given pressure, an enthalpy estimate function defined on a given structure $A$ is represented
as a linear sum over kernel functions
\begin{equation}
    H^\text{ML}(A) = \sum_{B \in M} w(B) K (A,B),
    \label{eqn:krr}
\end{equation}
which are summed over a representative set of structures $M$.  
The $M$ representative structures are selected from the total $N$ structures that are generated from RSS at each pressure.
For this selection, we exploited the farthest-point sampling (FPS) approach, which is a greedy algorithm that aims to select reference structures that are as diverse as possible by successively selecting a new point that is farthest from the ones already selected.
The set of weights $\{ w_{M} \}$ can be determined using
\begin{equation}
    w_M =
    (K_{MM}+K_{MN}\Lambda^{-1}K_{MN}^{\mathsf{T}})^{-1}
    K_{MN}\Lambda^{-1} H_N,
\end{equation}
where $K_{MM}$, $K_{NN}$, $K_{MN}$ denote the kernel matrix between the $M$ representative set, between the $N$ complete set, and between the representative and the complete set, respectively;
$\Lambda$ is an $N\times N$ diagonal matrix that is commonly used in the KRR framework to regularise the fit; and $H_N$ is the enthalpy per formula unit of the $N$ structures found by RSS.
To estimate the statistical error due to the finite size of the training set $N$,
we employed a subsampling technique~\cite{musil2019fast}: we created an ensemble of KRR models using a subset of the training data and used the variance of these model predictions to infer the uncertainties.
\rev{More details of the KRR model can be found in the \silabel{}.}

We plot the predictions of the KRR model alongside values obtained in direct energy minimisation simulations in the \silabel{} [Fig.~S2].
The agreement of the prediction with the calculated enthalpies is excellent, demonstrating that the ML model is able to capture the fundamentals of the interactions based on local environments, even though long-range Coulombic terms are present in the empirical potential.
\rev{The ML predictions for the basin entropy are less good, perhaps due to the intrinsic errors in the estimation of $s_\text{b}$ for systems of small sizes,
or because descriptions of equilibrium configurations cannot fully capture the basin volume.}
\rev{We note that representations other than the average SOAP kernel used here, as well as a different choice of the regression, can also be used for the predictions~\cite{faber2017prediction,Chen2019, Bartel2018}.}

\begin{figure*}
\includegraphics{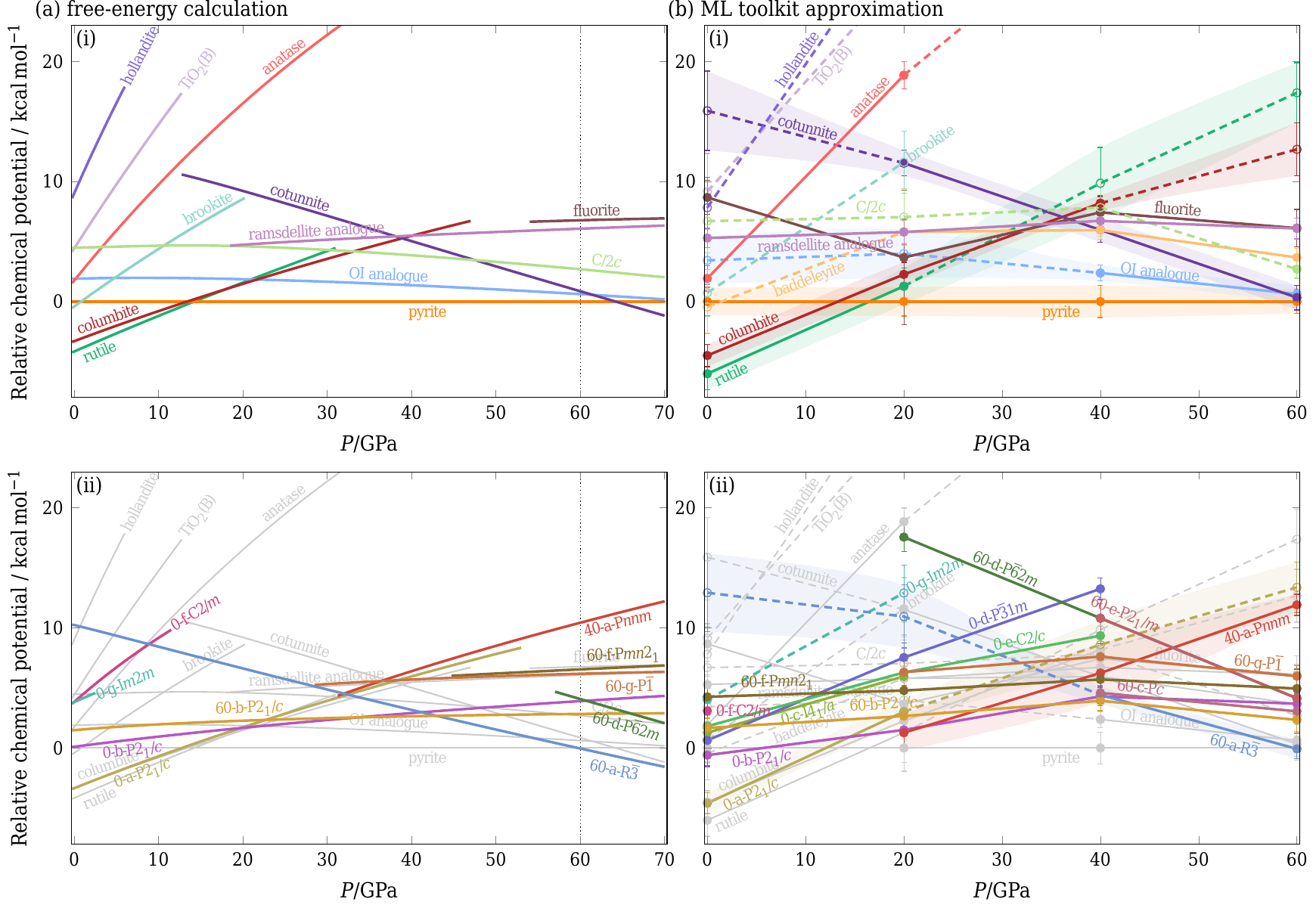}
\caption{
\rev{Chemical potential at \SI{1600}{\kelvin} expressed per formula unit and relative to the pyrite phase. 
Left panel: results obtained using free-energy calculations.
Right panel: predictions using solely data obtained from RSS. 
If a structure is found during RSS at a given pressure, solid lines show the approximate chemical potential based on its enthalpy and basin entropy. 
If a structure is not found, dashed lines indicate the machine-learning prediction as described in Section~\ref{subsect:ML-energetics}.}
In each case, in panel (i), only the previously known phases are shown, and in panel (ii), new phases obtained from RSS are shown, with the structures from panel (i) repeated in grey.
In (b), the error bar of potentially competitive phases is shown as a coloured band, and where a phase was not obtained at a given pressure from a random search, the marker is hollow and connected to other points by a dashed line. 
Note that the pressure axis covers different ranges in (a) and (b); dotted lines in (a) indicate the range of panel (b).}\label{fig:chemPot-MA-1600K}
\end{figure*}

\subsection{Phase behaviour and phase diagram estimation}

In \figrefsub{fig:chemPot-MA-1600K}{a}, we show chemical potentials (i.e.~Gibbs energies per formula unit) computed using TI at a relatively high temperature of \SI{1600}{\kelvin}.
We chose this temperature because 
entropic effects have become important, but the
majority of the phases of interest are still metastable across the pressure range considered.\footnote{\rev{It should however be noted that at high temperatures, intrinsic defects may become important~\cite{Nowotny2015,Sarkar2019}. Some defects, such as Frenkel-style interstitials, can be accounted for with our approach, but the most important defects in \titaniaForm{} include charge transfer and cannot be modelled with simple empirical potentials. Given the large typical free energies of defect formation of all main types even at such high temperatures~\cite{Nowotny2015}, it is not likely that they would significantly change the phase behaviour; however, an investigation of this behaviour with more advanced potentials would be particularly interesting.}}
The baddeleyite phase and several of the newly considered phases [namely 0-c-I4$_1$/\textit{a}, 0-e-C2/\textit{c}, 60-c-P\textit{c} and 60-e-P2$_1$/\textit{m}] do not appear in \figrefsub{fig:chemPot-MA-1600K}{a}, because they are no longer metastable at such high temperatures.
Baddeleyite, for example, spontaneously converts into pyrite at approximately \SI{1150}{\kelvin} when heated at \SI{20}{\giga\pascal}.
One particularly interesting transition is that between rutile and the new phase 40-a-P\textit{nnm} (see  \silabel{} (Fig.~S1)), which are structurally particularly similar.
The transition between them appears to be continuous as the system is pressurised: while the space group changes, there does not seem to be any change in density or enthalpy.
In a sense, the 40-a-P\textit{nnm} polymorph is thus merely a high-pressure analogue of the rutile phase.
Intriguingly, \figrefsub{fig:chemPot-MA-1600K}{a} shows a new phase, 60-a-R$\overline{3}$, is the most stable phase above $\sim$\SI{60}{\giga\pascal}, even though for some of this pressure regime, its enthalpy is not the lowest.
This example highlights the importance of properly accounting for the entropy of the polymorphs.
A number of other new phases also have free energies lower than some of the known structures of \titaniaForm{}, demonstrating the power of the RSS approach in successfully locating possible candidate phases for consideration.

To circumvent extensive and laborious free-energy calculations, we can approximate the chemical potential using solely the data obtained from RSS. 
If a structure is found in RSS at a particular pressure, 
we use $h(\SI{0}{\kelvin}) - Ts_\text{b}$ to approximate its chemical potential, where $h$ is the enthalpy per formula unit.
For structures that are not found, we first use the ML schemes outlined in Subsection~\ref{subsect:ML-energetics} to estimate the enthalpy and basin entropy, and then estimate the chemical potential using the same expression.
These chemical potentials are plotted in \figrefsub{fig:chemPot-MA-1600K}{b};
phases found in RSS are indicated by solid lines and phases not found by dashed lines.
As RSS was performed at four pressures, linear interpolations were used between the intervals.
Alternatively, if one has data at only one pressure,
a simple linear approximation $G(P+\upDelta P) \approx G(P) + \upDelta P V$ for extrapolation to nearby pressures can be considered~\cite{pickard2011ab}.

The agreement with the calculations of \figrefsub{fig:chemPot-MA-1600K}{a} is rather remarkable, with broad consistency of both the shape of the curves and the ordering of the structures (within error bars).
In part, this is because of the excellent prediction of the enthalpy (\silabel{}, Fig.~S2); in addition, we note that the approximate approach is also able to account for example for the fact that the OI phase has a low enthalpy, but a narrow basin of attraction, and becomes less favourable at sufficiently high temperatures.

We note that \figref{fig:chemPot-MA-1600K} 
encodes the information required for constructing phase diagrams:
the phase with the lowest chemical potential at a given pressure is the most stable for this potential at this temperature,
and the crossover of the chemical potential curves is a point of phase coexistence.
Using either the full free-energy calculation approach or the ML-based approximation, we can repeat the procedure at several temperatures and can thus determine the phase diagram of this system [see \silabel{}, Fig.~S7].
However, it ought to be borne in mind that a phase diagram hides much of the underlying thermodynamic behaviour, particularly for metastable phases,
so the prediction of chemical potentials, as illustrated in \figrefsub{fig:chemPot-MA-1600K}{b}, is thus a rather more rigorous test of the ML approach.
Furthermore, since the free energies of many of the phases considered are similar to one another and many phases are metastable under certain conditions, predicting the metastablities of all relevant polymorphs is particularly important.

\section{Conclusions}

\rev{
This paper presents a framework that uses a combination of ML methods to help extract information from random crystal structure searches.
This framework allows automatic classification and characterisation of possible solid phases,
as well as providing an estimate of the phase diagram and metastabilities of solid polymorphs. 
The framework does not use any prior knowledge of the phase behaviour of the system
and instead relies only on data obtained directly from a random structure search.
Starting from the thousands of structures obtained from RSS,
we first employ a pattern recognition approach to cluster similar structures with little manual input.
We then estimate the basin entropy of the crystal structures found by assuming that their basin of attraction is related to the number of times each structure is found.
Moreover, we use a machine learning approach to predict the enthalpy of unknown structures based on ones that are observed.
Combining all these, we can estimate the free energies of solid phases and hence the thermodynamic phase behaviour of a system of interest with relatively little computational effort.
}

We tested the framework on the rather complicated system of \titaniaForm{} at ambient and high pressures.
We predicted a number of possible phases and estimated their chemical potentials at finite temperatures. 
Even though we only explicitly showed the chemical potentials at a moderately high temperature,
the framework that we present can be used for any reasonable temperature.
Pair potentials similar to the ones used for our RSS are widely used for other binary oxides~\cite{Lewis1985, Shimojo1992},
so the new phases we found and the methodology introduced may be relevant to other oxides as well.

The methods we propose are transferable to other systems with many solid polymorphs, such as high-pressure solid hydrogen~\cite{cheng2019evidence},
perovskites~\cite{murakami2004post}, different phases of water-ice~\cite{engel2018mapping} and molecular crystals~\cite{beran2016modeling}.
\rev{A useful strategy for predicting the phase diagram is first to use the framework to select competitive polymorphs using the data generated from crystal structure searches,
and then to compute explicit free energies \emph{only} of those phases.}
Such an approach would make the determination of phase behaviour of systems exhibiting many solid polymorphs using accurate free-energy calculations considerably more straightforward and computationally tractable.

\section*{Acknowledgements}
We acknowledge fruitful discussions with Daan Frenkel.
We thank Michele Ceriotti, Stephen J.~Cox, Edgar Engel and Bonan Zhu for critically reading the manuscript and providing useful feedback. 
BC acknowledges funding from the Swiss National Science Foundation (Project P2ELP2-184408).
AR and BC acknowledge resources provided by the Cambridge Tier-2 system operated by the University of Cambridge Research Computing Service funded by EPSRC Tier-2 capital grant EP/P020259/1.
CJP acknowledges financial support from the Engineering and Physical Sciences Research Council [Grant EP/P022596/1] and a Royal Society Wolfson Research Merit award.

The authors declare no competing interests.


\begin{thebibliography}{61}%
\makeatletter
\providecommand \@ifxundefined [1]{%
 \@ifx{#1\undefined}
}%
\providecommand \@ifnum [1]{%
 \ifnum #1\expandafter \@firstoftwo
 \else \expandafter \@secondoftwo
 \fi
}%
\providecommand \@ifx [1]{%
 \ifx #1\expandafter \@firstoftwo
 \else \expandafter \@secondoftwo
 \fi
}%
\providecommand \natexlab [1]{#1}%
\providecommand \enquote  [1]{``#1''}%
\providecommand \bibnamefont  [1]{#1}%
\providecommand \bibfnamefont [1]{#1}%
\providecommand \citenamefont [1]{#1}%
\providecommand \href@noop [0]{\@secondoftwo}%
\providecommand \href [0]{\begingroup \@sanitize@url \@href}%
\providecommand \@href[1]{\@@startlink{#1}\@@href}%
\providecommand \@@href[1]{\endgroup#1\@@endlink}%
\providecommand \@sanitize@url [0]{\catcode `\\12\catcode `\$12\catcode
  `\&12\catcode `\#12\catcode `\^12\catcode `\_12\catcode `\%12\relax}%
\providecommand \@@startlink[1]{}%
\providecommand \@@endlink[0]{}%
\providecommand \url  [0]{\begingroup\@sanitize@url \@url }%
\providecommand \@url [1]{\endgroup\@href {#1}{\urlprefix }}%
\providecommand \urlprefix  [0]{URL }%
\providecommand \Eprint [0]{\href }%
\providecommand \doibase [0]{http://dx.doi.org/}%
\providecommand \selectlanguage [0]{\@gobble}%
\providecommand \bibinfo  [0]{\@secondoftwo}%
\providecommand \bibfield  [0]{\@secondoftwo}%
\providecommand \translation [1]{[#1]}%
\providecommand \BibitemOpen [0]{}%
\providecommand \bibitemStop [0]{}%
\providecommand \bibitemNoStop [0]{.\EOS\space}%
\providecommand \EOS [0]{\spacefactor3000\relax}%
\providecommand \BibitemShut  [1]{\csname bibitem#1\endcsname}%
\let\auto@bib@innerbib\@empty
\bibitem [{\citenamefont {Woodley}\ and\ \citenamefont
  {Catlow}(2008)}]{woodley2008crystal}%
  \BibitemOpen
  \bibfield  {author} {\bibinfo {author} {\bibfnamefont {S.~M.}\ \bibnamefont
  {Woodley}}\ and\ \bibinfo {author} {\bibfnamefont {R.}~\bibnamefont
  {Catlow}},\ }\bibfield  {title} {\enquote {\bibinfo {title} {Crystal
  structure prediction from first principles},}\ }\href {\doibase
  10.1038/nmat2321} {\bibfield  {journal} {\bibinfo  {journal} {Nat.\ Mater.}\
  }\textbf {\bibinfo {volume} {7}},\ \bibinfo {pages} {937--946} (\bibinfo
  {year} {2008})}\BibitemShut {NoStop}%
\bibitem [{\citenamefont {Pickard}\ and\ \citenamefont
  {Needs}(2011{\natexlab{a}})}]{pickard2011ab}%
  \BibitemOpen
  \bibfield  {author} {\bibinfo {author} {\bibfnamefont {C.~J.}\ \bibnamefont
  {Pickard}}\ and\ \bibinfo {author} {\bibfnamefont {R.}~\bibnamefont
  {Needs}},\ }\bibfield  {title} {\enquote {\bibinfo {title} {\textit{Ab
  initio} random structure searching},}\ }\href {\doibase
  10.1088/0953-8984/23/5/053201} {\bibfield  {journal} {\bibinfo  {journal}
  {J.\ Phys.: Condens.\ Matter}\ }\textbf {\bibinfo {volume} {23}},\ \bibinfo
  {pages} {053201} (\bibinfo {year} {2011}{\natexlab{a}})}\BibitemShut
  {NoStop}%
\bibitem [{\citenamefont {Wang}\ \emph {et~al.}(2010)\citenamefont {Wang},
  \citenamefont {Lv}, \citenamefont {Zhu},\ and\ \citenamefont
  {Ma}}]{wang2010crystal}%
  \BibitemOpen
  \bibfield  {author} {\bibinfo {author} {\bibfnamefont {Y.}~\bibnamefont
  {Wang}}, \bibinfo {author} {\bibfnamefont {J.}~\bibnamefont {Lv}}, \bibinfo
  {author} {\bibfnamefont {L.}~\bibnamefont {Zhu}}, \ and\ \bibinfo {author}
  {\bibfnamefont {Y.}~\bibnamefont {Ma}},\ }\bibfield  {title} {\enquote
  {\bibinfo {title} {Crystal structure prediction via particle-swarm
  optimization},}\ }\href {\doibase 10.1103/PhysRevB.82.094116} {\bibfield
  {journal} {\bibinfo  {journal} {Phys.\ Rev.\ B}\ }\textbf {\bibinfo {volume}
  {82}},\ \bibinfo {pages} {094116} (\bibinfo {year} {2010})}\BibitemShut
  {NoStop}%
\bibitem [{\citenamefont {Filion}\ \emph {et~al.}(2009)\citenamefont {Filion},
  \citenamefont {Marechal}, \citenamefont {van Oorschot}, \citenamefont {Pelt},
  \citenamefont {Smallenburg},\ and\ \citenamefont
  {Dijkstra}}]{filion2009efficient}%
  \BibitemOpen
  \bibfield  {author} {\bibinfo {author} {\bibfnamefont {L.}~\bibnamefont
  {Filion}}, \bibinfo {author} {\bibfnamefont {M.}~\bibnamefont {Marechal}},
  \bibinfo {author} {\bibfnamefont {B.}~\bibnamefont {van Oorschot}}, \bibinfo
  {author} {\bibfnamefont {D.}~\bibnamefont {Pelt}}, \bibinfo {author}
  {\bibfnamefont {F.}~\bibnamefont {Smallenburg}}, \ and\ \bibinfo {author}
  {\bibfnamefont {M.}~\bibnamefont {Dijkstra}},\ }\bibfield  {title} {\enquote
  {\bibinfo {title} {Efficient method for predicting crystal structures at
  finite temperature: {V}ariable box shape simulations},}\ }\href {\doibase
  10.1103/PhysRevLett.103.188302} {\bibfield  {journal} {\bibinfo  {journal}
  {Phys.\ Rev.\ Lett.}\ }\textbf {\bibinfo {volume} {103}},\ \bibinfo {pages}
  {188302} (\bibinfo {year} {2009})}\BibitemShut {NoStop}%
\bibitem [{\citenamefont {Wales}\ and\ \citenamefont
  {Scheraga}(1999)}]{wales1999global}%
  \BibitemOpen
  \bibfield  {author} {\bibinfo {author} {\bibfnamefont {D.~J.}\ \bibnamefont
  {Wales}}\ and\ \bibinfo {author} {\bibfnamefont {H.~A.}\ \bibnamefont
  {Scheraga}},\ }\bibfield  {title} {\enquote {\bibinfo {title} {Global
  optimization of clusters, crystals, and biomolecules},}\ }\href {\doibase
  10.1126/science.285.5432.1368} {\bibfield  {journal} {\bibinfo  {journal}
  {Science}\ }\textbf {\bibinfo {volume} {285}},\ \bibinfo {pages} {1368--1372}
  (\bibinfo {year} {1999})}\BibitemShut {NoStop}%
\bibitem [{\citenamefont {Cheng}\ and\ \citenamefont
  {Ceriotti}(2018)}]{cheng2018computing}%
  \BibitemOpen
  \bibfield  {author} {\bibinfo {author} {\bibfnamefont {B.}~\bibnamefont
  {Cheng}}\ and\ \bibinfo {author} {\bibfnamefont {M.}~\bibnamefont
  {Ceriotti}},\ }\bibfield  {title} {\enquote {\bibinfo {title} {Computing the
  absolute {G}ibbs free energy in atomistic simulations: {A}pplications to
  defects in solids},}\ }\href {\doibase 10.1103/PhysRevB.97.054102} {\bibfield
   {journal} {\bibinfo  {journal} {Phys.\ Rev.\ B}\ }\textbf {\bibinfo {volume}
  {97}},\ \bibinfo {pages} {054102} (\bibinfo {year} {2018})}\BibitemShut
  {NoStop}%
\bibitem [{\citenamefont {Fu}\ \emph {et~al.}(2013)\citenamefont {Fu},
  \citenamefont {Liang}, \citenamefont {Wang},\ and\ \citenamefont
  {Zhong}}]{Fu2013}%
  \BibitemOpen
  \bibfield  {author} {\bibinfo {author} {\bibfnamefont {Z.}~\bibnamefont
  {Fu}}, \bibinfo {author} {\bibfnamefont {Y.}~\bibnamefont {Liang}}, \bibinfo
  {author} {\bibfnamefont {S.}~\bibnamefont {Wang}}, \ and\ \bibinfo {author}
  {\bibfnamefont {Z.}~\bibnamefont {Zhong}},\ }\bibfield  {title} {\enquote
  {\bibinfo {title} {Structural phase transition and mechanical properties of
  {{\ce{TiO2}}} under high pressure},}\ }\href {\doibase
  10.1002/pssb.201349186} {\bibfield  {journal} {\bibinfo  {journal} {Phys.\
  Status Solidi B}\ }\textbf {\bibinfo {volume} {250}},\ \bibinfo {pages}
  {2206--2214} (\bibinfo {year} {2013})}\BibitemShut {NoStop}%
\bibitem [{\citenamefont {Vega}\ \emph {et~al.}(2008)\citenamefont {Vega},
  \citenamefont {Sanz}, \citenamefont {Abascal},\ and\ \citenamefont
  {Noya}}]{Vega2008}%
  \BibitemOpen
  \bibfield  {author} {\bibinfo {author} {\bibfnamefont {C.}~\bibnamefont
  {Vega}}, \bibinfo {author} {\bibfnamefont {E.}~\bibnamefont {Sanz}}, \bibinfo
  {author} {\bibfnamefont {J.~L.~F.}\ \bibnamefont {Abascal}}, \ and\ \bibinfo
  {author} {\bibfnamefont {E.~G.}\ \bibnamefont {Noya}},\ }\bibfield  {title}
  {\enquote {\bibinfo {title} {Determination of phase diagrams via computer
  simulation: {M}ethodology and applications to water, electrolytes and
  proteins},}\ }\href {\doibase 10.1088/0953-8984/20/15/153101} {\bibfield
  {journal} {\bibinfo  {journal} {J.\ Phys.:\ Condens.\ Matter}\ }\textbf
  {\bibinfo {volume} {20}},\ \bibinfo {pages} {153101} (\bibinfo {year}
  {2008})}\BibitemShut {NoStop}%
\bibitem [{\citenamefont {Reinhardt}(2019)}]{Reinhardt2019}%
  \BibitemOpen
  \bibfield  {author} {\bibinfo {author} {\bibfnamefont {A.}~\bibnamefont
  {Reinhardt}},\ }\bibfield  {title} {\enquote {\bibinfo {title} {Phase
  behavior of empirical potentials of titanium dioxide},}\ }\href {\doibase
  10.1063/1.5115161} {\bibfield  {journal} {\bibinfo  {journal} {J.\ Chem.\
  Phys.}\ }\textbf {\bibinfo {volume} {151}},\ \bibinfo {pages} {064505}
  (\bibinfo {year} {2019})}\BibitemShut {NoStop}%
\bibitem [{\citenamefont {Monserrat}\ \emph {et~al.}(2018)\citenamefont
  {Monserrat}, \citenamefont {Drummond}, \citenamefont {Dalladay-Simpson},
  \citenamefont {Howie}, \citenamefont {R{\'\i}os}, \citenamefont {Gregoryanz},
  \citenamefont {Pickard},\ and\ \citenamefont
  {Needs}}]{monserrat2018structure}%
  \BibitemOpen
  \bibfield  {author} {\bibinfo {author} {\bibfnamefont {B.}~\bibnamefont
  {Monserrat}}, \bibinfo {author} {\bibfnamefont {N.~D.}\ \bibnamefont
  {Drummond}}, \bibinfo {author} {\bibfnamefont {P.}~\bibnamefont
  {Dalladay-Simpson}}, \bibinfo {author} {\bibfnamefont {R.~T.}\ \bibnamefont
  {Howie}}, \bibinfo {author} {\bibfnamefont {P.~L.}\ \bibnamefont
  {R{\'\i}os}}, \bibinfo {author} {\bibfnamefont {E.}~\bibnamefont
  {Gregoryanz}}, \bibinfo {author} {\bibfnamefont {C.~J.}\ \bibnamefont
  {Pickard}}, \ and\ \bibinfo {author} {\bibfnamefont {R.~J.}\ \bibnamefont
  {Needs}},\ }\bibfield  {title} {\enquote {\bibinfo {title} {Structure and
  metallicity of phase {V} of hydrogen},}\ }\href {\doibase
  10.1103/PhysRevLett.120.255701} {\bibfield  {journal} {\bibinfo  {journal}
  {Phys.\ Rev.\ Lett.}\ }\textbf {\bibinfo {volume} {120}},\ \bibinfo {pages}
  {255701} (\bibinfo {year} {2018})}\BibitemShut {NoStop}%
\bibitem [{\citenamefont {Dekura}\ \emph {et~al.}(2011)\citenamefont {Dekura},
  \citenamefont {Tsuchiya}, \citenamefont {Kuwayama},\ and\ \citenamefont
  {Tsuchiya}}]{Dekura2011}%
  \BibitemOpen
  \bibfield  {author} {\bibinfo {author} {\bibfnamefont {H.}~\bibnamefont
  {Dekura}}, \bibinfo {author} {\bibfnamefont {T.}~\bibnamefont {Tsuchiya}},
  \bibinfo {author} {\bibfnamefont {Y.}~\bibnamefont {Kuwayama}}, \ and\
  \bibinfo {author} {\bibfnamefont {J.}~\bibnamefont {Tsuchiya}},\ }\bibfield
  {title} {\enquote {\bibinfo {title} {Theoretical and experimental evidence
  for a new post-cotunnite phase of titanium dioxide with significant optical
  absorption},}\ }\href {\doibase 10.1103/PhysRevLett.107.045701} {\bibfield
  {journal} {\bibinfo  {journal} {Phys.\ Rev.\ Lett.}\ }\textbf {\bibinfo
  {volume} {107}},\ \bibinfo {pages} {045701} (\bibinfo {year}
  {2011})}\BibitemShut {NoStop}%
\bibitem [{\citenamefont {Dubrovinsky}\ \emph {et~al.}(2001)\citenamefont
  {Dubrovinsky}, \citenamefont {Dubrovinskaia}, \citenamefont {Swamy},
  \citenamefont {Muscat}, \citenamefont {Harrison}, \citenamefont {Ahuja},
  \citenamefont {Holm},\ and\ \citenamefont {Johansson}}]{Dubrovinsky2001}%
  \BibitemOpen
  \bibfield  {author} {\bibinfo {author} {\bibfnamefont {L.~S.}\ \bibnamefont
  {Dubrovinsky}}, \bibinfo {author} {\bibfnamefont {N.~A.}\ \bibnamefont
  {Dubrovinskaia}}, \bibinfo {author} {\bibfnamefont {V.}~\bibnamefont
  {Swamy}}, \bibinfo {author} {\bibfnamefont {J.}~\bibnamefont {Muscat}},
  \bibinfo {author} {\bibfnamefont {N.~M.}\ \bibnamefont {Harrison}}, \bibinfo
  {author} {\bibfnamefont {R.}~\bibnamefont {Ahuja}}, \bibinfo {author}
  {\bibfnamefont {B.}~\bibnamefont {Holm}}, \ and\ \bibinfo {author}
  {\bibfnamefont {B.}~\bibnamefont {Johansson}},\ }\bibfield  {title} {\enquote
  {\bibinfo {title} {The hardest known oxide},}\ }\href {\doibase
  10.1038/35070650} {\bibfield  {journal} {\bibinfo  {journal} {Nature}\
  }\textbf {\bibinfo {volume} {410}},\ \bibinfo {pages} {653--654} (\bibinfo
  {year} {2001})}\BibitemShut {NoStop}%
\bibitem [{\citenamefont {Mukai}\ and\ \citenamefont
  {Yamada}(2017)}]{Mukai2017}%
  \BibitemOpen
  \bibfield  {author} {\bibinfo {author} {\bibfnamefont {K.}~\bibnamefont
  {Mukai}}\ and\ \bibinfo {author} {\bibfnamefont {I.}~\bibnamefont {Yamada}},\
  }\bibfield  {title} {\enquote {\bibinfo {title} {Columbite-type \ce{TiO2} as
  a negative electrode material for lithium-ion batteries},}\ }\href {\doibase
  10.1149/2.0481714jes} {\bibfield  {journal} {\bibinfo  {journal} {J.\
  Electrochem.\ Soc.}\ }\textbf {\bibinfo {volume} {164}},\ \bibinfo {pages}
  {A3590--A3594} (\bibinfo {year} {2017})}\BibitemShut {NoStop}%
\bibitem [{\citenamefont {{Staun Olsen}}\ \emph {et~al.}(1999)\citenamefont
  {{Staun Olsen}}, \citenamefont {Gerward},\ and\ \citenamefont
  {Jiang}}]{StaunOlsen1999}%
  \BibitemOpen
  \bibfield  {author} {\bibinfo {author} {\bibfnamefont {J.}~\bibnamefont
  {{Staun Olsen}}}, \bibinfo {author} {\bibfnamefont {L.}~\bibnamefont
  {Gerward}}, \ and\ \bibinfo {author} {\bibfnamefont {J.~Z.}\ \bibnamefont
  {Jiang}},\ }\bibfield  {title} {\enquote {\bibinfo {title} {On the
  rutile/$\upalpha$-\ce{PbO2}-type phase boundary of \ce{TiO2}},}\ }\href
  {\doibase 10.1016/S0022-3697(98)00274-1} {\bibfield  {journal} {\bibinfo
  {journal} {J.\ Phys.\ Chem.\ Solids}\ }\textbf {\bibinfo {volume} {60}},\
  \bibinfo {pages} {229--233} (\bibinfo {year} {1999})}\BibitemShut {NoStop}%
\bibitem [{\citenamefont {Matsui}\ and\ \citenamefont
  {Akaogi}(1991)}]{Matsui1991}%
  \BibitemOpen
  \bibfield  {author} {\bibinfo {author} {\bibfnamefont {M.}~\bibnamefont
  {Matsui}}\ and\ \bibinfo {author} {\bibfnamefont {M.}~\bibnamefont
  {Akaogi}},\ }\bibfield  {title} {\enquote {\bibinfo {title} {Molecular
  dynamics simulation of the structural and physical properties of the four
  polymorphs of {{\ce{TiO2}}}},}\ }\href {\doibase 10.1080/08927029108022432}
  {\bibfield  {journal} {\bibinfo  {journal} {Mol.\ Simul.}\ }\textbf {\bibinfo
  {volume} {6}},\ \bibinfo {pages} {239--244} (\bibinfo {year}
  {1991})}\BibitemShut {NoStop}%
\bibitem [{\citenamefont {Pickard}\ and\ \citenamefont
  {Needs}(2006)}]{Pickard2006}%
  \BibitemOpen
  \bibfield  {author} {\bibinfo {author} {\bibfnamefont {C.~J.}\ \bibnamefont
  {Pickard}}\ and\ \bibinfo {author} {\bibfnamefont {R.~J.}\ \bibnamefont
  {Needs}},\ }\bibfield  {title} {\enquote {\bibinfo {title} {High-pressure
  phases of silane},}\ }\href {\doibase 10.1103/PhysRevLett.97.045504}
  {\bibfield  {journal} {\bibinfo  {journal} {Phys.\ Rev.\ Lett.}\ }\textbf
  {\bibinfo {volume} {97}},\ \bibinfo {pages} {045504} (\bibinfo {year}
  {2006})}\BibitemShut {NoStop}%
\bibitem [{\citenamefont {Pickard}\ and\ \citenamefont
  {Needs}(2011{\natexlab{b}})}]{Pickard2011}%
  \BibitemOpen
  \bibfield  {author} {\bibinfo {author} {\bibfnamefont {C.~J.}\ \bibnamefont
  {Pickard}}\ and\ \bibinfo {author} {\bibfnamefont {R.~J.}\ \bibnamefont
  {Needs}},\ }\bibfield  {title} {\enquote {\bibinfo {title} {\textit{Ab
  initio} random structure searching},}\ }\href {\doibase
  10.1088/0953-8984/23/5/053201} {\bibfield  {journal} {\bibinfo  {journal}
  {J.\ Phys.:\ Condens.\ Matter}\ }\textbf {\bibinfo {volume} {23}},\ \bibinfo
  {pages} {053201} (\bibinfo {year} {2011}{\natexlab{b}})}\BibitemShut
  {NoStop}%
\bibitem [{\citenamefont {Plimpton}(1995)}]{Plimpton1995}%
  \BibitemOpen
  \bibfield  {author} {\bibinfo {author} {\bibfnamefont {S.}~\bibnamefont
  {Plimpton}},\ }\bibfield  {title} {\enquote {\bibinfo {title} {Fast parallel
  algorithms for short-range molecular dynamics},}\ }\href {\doibase
  10.1006/jcph.1995.1039} {\bibfield  {journal} {\bibinfo  {journal} {J.\
  Comput.\ Phys.}\ }\textbf {\bibinfo {volume} {117}},\ \bibinfo {pages}
  {1--19} (\bibinfo {year} {1995})}\BibitemShut {NoStop}%
\bibitem [{\citenamefont {Frenkel}\ and\ \citenamefont
  {Ladd}(1984)}]{Frenkel1984}%
  \BibitemOpen
  \bibfield  {author} {\bibinfo {author} {\bibfnamefont {D.}~\bibnamefont
  {Frenkel}}\ and\ \bibinfo {author} {\bibfnamefont {A.~J.~C.}\ \bibnamefont
  {Ladd}},\ }\bibfield  {title} {\enquote {\bibinfo {title} {New {M}onte
  {C}arlo method to compute the free energy of arbitrary solids. {A}pplication
  to the fcc and hcp phases of hard spheres},}\ }\href {\doibase
  10.1063/1.448024} {\bibfield  {journal} {\bibinfo  {journal} {J.\ Chem.\
  Phys.}\ }\textbf {\bibinfo {volume} {81}},\ \bibinfo {pages} {3188--3193}
  (\bibinfo {year} {1984})}\BibitemShut {NoStop}%
\bibitem [{\citenamefont {Jones}\ and\ \citenamefont
  {Gunnarsson}(1989)}]{jones1989density}%
  \BibitemOpen
  \bibfield  {author} {\bibinfo {author} {\bibfnamefont {R.~O.}\ \bibnamefont
  {Jones}}\ and\ \bibinfo {author} {\bibfnamefont {O.}~\bibnamefont
  {Gunnarsson}},\ }\bibfield  {title} {\enquote {\bibinfo {title} {The density
  functional formalism, its applications and prospects},}\ }\href {\doibase
  10.1103/RevModPhys.61.689} {\bibfield  {journal} {\bibinfo  {journal} {Rev.\
  Mod.\ Phys.}\ }\textbf {\bibinfo {volume} {61}},\ \bibinfo {pages} {689--746}
  (\bibinfo {year} {1989})}\BibitemShut {NoStop}%
\bibitem [{\citenamefont {Perdew}\ \emph {et~al.}(1996)\citenamefont {Perdew},
  \citenamefont {Burke},\ and\ \citenamefont
  {Ernzerhof}}]{perdew1996generalized}%
  \BibitemOpen
  \bibfield  {author} {\bibinfo {author} {\bibfnamefont {J.~P.}\ \bibnamefont
  {Perdew}}, \bibinfo {author} {\bibfnamefont {K.}~\bibnamefont {Burke}}, \
  and\ \bibinfo {author} {\bibfnamefont {M.}~\bibnamefont {Ernzerhof}},\
  }\bibfield  {title} {\enquote {\bibinfo {title} {Generalized gradient
  approximation made simple},}\ }\href {\doibase 10.1103/PhysRevLett.77.3865}
  {\bibfield  {journal} {\bibinfo  {journal} {Phys.\ Rev.\ Lett.}\ }\textbf
  {\bibinfo {volume} {77}},\ \bibinfo {pages} {3865--3868} (\bibinfo {year}
  {1996})}\BibitemShut {NoStop}%
\bibitem [{\citenamefont {Perdew}\ \emph {et~al.}(2008)\citenamefont {Perdew},
  \citenamefont {Ruzsinszky}, \citenamefont {Csonka}, \citenamefont {Vydrov},
  \citenamefont {Scuseria}, \citenamefont {Constantin}, \citenamefont {Zhou},\
  and\ \citenamefont {Burke}}]{perdew2008restoring}%
  \BibitemOpen
  \bibfield  {author} {\bibinfo {author} {\bibfnamefont {J.~P.}\ \bibnamefont
  {Perdew}}, \bibinfo {author} {\bibfnamefont {A.}~\bibnamefont {Ruzsinszky}},
  \bibinfo {author} {\bibfnamefont {G.~I.}\ \bibnamefont {Csonka}}, \bibinfo
  {author} {\bibfnamefont {O.~A.}\ \bibnamefont {Vydrov}}, \bibinfo {author}
  {\bibfnamefont {G.~E.}\ \bibnamefont {Scuseria}}, \bibinfo {author}
  {\bibfnamefont {L.~A.}\ \bibnamefont {Constantin}}, \bibinfo {author}
  {\bibfnamefont {X.}~\bibnamefont {Zhou}}, \ and\ \bibinfo {author}
  {\bibfnamefont {K.}~\bibnamefont {Burke}},\ }\bibfield  {title} {\enquote
  {\bibinfo {title} {Restoring the density-gradient expansion for exchange in
  solids and surfaces},}\ }\href {\doibase 10.1103/PhysRevLett.100.136406}
  {\bibfield  {journal} {\bibinfo  {journal} {Phys.\ Rev.\ Lett.}\ }\textbf
  {\bibinfo {volume} {100}},\ \bibinfo {pages} {136406} (\bibinfo {year}
  {2008})}\BibitemShut {NoStop}%
\bibitem [{\citenamefont {Clark}\ \emph {et~al.}(2005)\citenamefont {Clark},
  \citenamefont {Segall}, \citenamefont {Pickard}, \citenamefont {Hasnip},
  \citenamefont {Probert}, \citenamefont {Refson},\ and\ \citenamefont
  {Payne}}]{clark2005first}%
  \BibitemOpen
  \bibfield  {author} {\bibinfo {author} {\bibfnamefont {S.~J.}\ \bibnamefont
  {Clark}}, \bibinfo {author} {\bibfnamefont {M.~D.}\ \bibnamefont {Segall}},
  \bibinfo {author} {\bibfnamefont {C.~J.}\ \bibnamefont {Pickard}}, \bibinfo
  {author} {\bibfnamefont {P.~J.}\ \bibnamefont {Hasnip}}, \bibinfo {author}
  {\bibfnamefont {M.~I.}\ \bibnamefont {Probert}}, \bibinfo {author}
  {\bibfnamefont {K.}~\bibnamefont {Refson}}, \ and\ \bibinfo {author}
  {\bibfnamefont {M.~C.}\ \bibnamefont {Payne}},\ }\bibfield  {title} {\enquote
  {\bibinfo {title} {First principles methods using {CASTEP}},}\ }\href
  {\doibase 10.1524/zkri.220.5.567.65075} {\bibfield  {journal} {\bibinfo
  {journal} {Z.\ Kristallogr.\ Cryst.\ Mater.}\ }\textbf {\bibinfo {volume}
  {220}},\ \bibinfo {pages} {567--570} (\bibinfo {year} {2005})}\BibitemShut
  {NoStop}%
\bibitem [{\citenamefont {Ceriotti}\ \emph {et~al.}(2011)\citenamefont
  {Ceriotti}, \citenamefont {Tribello},\ and\ \citenamefont
  {Parrinello}}]{ceriotti2011simplifying}%
  \BibitemOpen
  \bibfield  {author} {\bibinfo {author} {\bibfnamefont {M.}~\bibnamefont
  {Ceriotti}}, \bibinfo {author} {\bibfnamefont {G.~A.}\ \bibnamefont
  {Tribello}}, \ and\ \bibinfo {author} {\bibfnamefont {M.}~\bibnamefont
  {Parrinello}},\ }\bibfield  {title} {\enquote {\bibinfo {title} {Simplifying
  the representation of complex free-energy landscapes using sketch-map},}\
  }\href {\doibase 10.1073/pnas.1108486108} {\bibfield  {journal} {\bibinfo
  {journal} {Proc.\ Natl.\ Acad.\ Sci.\ U.~S.~A.}\ }\textbf {\bibinfo {volume}
  {108}},\ \bibinfo {pages} {13023--13028} (\bibinfo {year}
  {2011})}\BibitemShut {NoStop}%
\bibitem [{\citenamefont {Engel}\ \emph {et~al.}(2018)\citenamefont {Engel},
  \citenamefont {Anelli}, \citenamefont {Ceriotti}, \citenamefont {Pickard},\
  and\ \citenamefont {Needs}}]{engel2018mapping}%
  \BibitemOpen
  \bibfield  {author} {\bibinfo {author} {\bibfnamefont {E.~A.}\ \bibnamefont
  {Engel}}, \bibinfo {author} {\bibfnamefont {A.}~\bibnamefont {Anelli}},
  \bibinfo {author} {\bibfnamefont {M.}~\bibnamefont {Ceriotti}}, \bibinfo
  {author} {\bibfnamefont {C.~J.}\ \bibnamefont {Pickard}}, \ and\ \bibinfo
  {author} {\bibfnamefont {R.~J.}\ \bibnamefont {Needs}},\ }\bibfield  {title}
  {\enquote {\bibinfo {title} {Mapping uncharted territory in ice from zeolite
  networks to ice structures},}\ }\href {\doibase 10.1038/s41467-018-04618-6}
  {\bibfield  {journal} {\bibinfo  {journal} {Nat.\ Commun.}\ }\textbf
  {\bibinfo {volume} {9}},\ \bibinfo {pages} {2173} (\bibinfo {year}
  {2018})}\BibitemShut {NoStop}%
\bibitem [{\citenamefont {De}\ \emph {et~al.}(2016)\citenamefont {De},
  \citenamefont {Bart{\'{o}}k}, \citenamefont {Cs{\'{a}}nyi},\ and\
  \citenamefont {Ceriotti}}]{de+16pccp}%
  \BibitemOpen
  \bibfield  {author} {\bibinfo {author} {\bibfnamefont {S.}~\bibnamefont
  {De}}, \bibinfo {author} {\bibfnamefont {A.~P.}\ \bibnamefont
  {Bart{\'{o}}k}}, \bibinfo {author} {\bibfnamefont {G.}~\bibnamefont
  {Cs{\'{a}}nyi}}, \ and\ \bibinfo {author} {\bibfnamefont {M.}~\bibnamefont
  {Ceriotti}},\ }\bibfield  {title} {\enquote {\bibinfo {title} {Comparing
  molecules and solids across structural and alchemical space},}\ }\href
  {\doibase 10.1039/c6cp00415f} {\bibfield  {journal} {\bibinfo  {journal}
  {Phys.\ Chem.\ Chem.\ Phys.}\ }\textbf {\bibinfo {volume} {18}},\ \bibinfo
  {pages} {13754--13769} (\bibinfo {year} {2016})}\BibitemShut {NoStop}%
\bibitem [{\citenamefont {Anelli}\ \emph {et~al.}(2018)\citenamefont {Anelli},
  \citenamefont {Engel}, \citenamefont {Pickard},\ and\ \citenamefont
  {Ceriotti}}]{anelli2018generalized}%
  \BibitemOpen
  \bibfield  {author} {\bibinfo {author} {\bibfnamefont {A.}~\bibnamefont
  {Anelli}}, \bibinfo {author} {\bibfnamefont {E.~A.}\ \bibnamefont {Engel}},
  \bibinfo {author} {\bibfnamefont {C.~J.}\ \bibnamefont {Pickard}}, \ and\
  \bibinfo {author} {\bibfnamefont {M.}~\bibnamefont {Ceriotti}},\ }\bibfield
  {title} {\enquote {\bibinfo {title} {Generalized convex hull construction for
  materials discovery},}\ }\href {\doibase 10.1103/PhysRevMaterials.2.103804}
  {\bibfield  {journal} {\bibinfo  {journal} {Phys.\ Rev.\ Mater.}\ }\textbf
  {\bibinfo {volume} {2}},\ \bibinfo {pages} {103804} (\bibinfo {year}
  {2018})}\BibitemShut {NoStop}%
\bibitem [{\citenamefont {Mavra\v{c}i\'{c}}\ \emph {et~al.}(2018)\citenamefont
  {Mavra\v{c}i\'{c}}, \citenamefont {Mocanu}, \citenamefont {Deringer},
  \citenamefont {Cs\'{a}nyi},\ and\ \citenamefont {Elliott}}]{Mavracic2018}%
  \BibitemOpen
  \bibfield  {author} {\bibinfo {author} {\bibfnamefont {J.}~\bibnamefont
  {Mavra\v{c}i\'{c}}}, \bibinfo {author} {\bibfnamefont {F.~C.}\ \bibnamefont
  {Mocanu}}, \bibinfo {author} {\bibfnamefont {V.~L.}\ \bibnamefont
  {Deringer}}, \bibinfo {author} {\bibfnamefont {G.}~\bibnamefont
  {Cs\'{a}nyi}}, \ and\ \bibinfo {author} {\bibfnamefont {S.~R.}\ \bibnamefont
  {Elliott}},\ }\bibfield  {title} {\enquote {\bibinfo {title} {Similarity
  between amorphous and crystalline phases: The case of \ce{TiO2}},}\ }\href
  {\doibase 10.1021/acs.jpclett.8b01067} {\bibfield  {journal} {\bibinfo
  {journal} {J.\ Phys.\ Chem.\ Lett.}\ }\textbf {\bibinfo {volume} {9}},\
  \bibinfo {pages} {2985--2990} (\bibinfo {year} {2018})}\BibitemShut {NoStop}%
\bibitem [{\citenamefont {Bart{\'{o}}k}\ \emph {et~al.}(2013)\citenamefont
  {Bart{\'{o}}k}, \citenamefont {Kondor},\ and\ \citenamefont
  {Cs{\'{a}}nyi}}]{bart+13prb}%
  \BibitemOpen
  \bibfield  {author} {\bibinfo {author} {\bibfnamefont {A.~P.}\ \bibnamefont
  {Bart{\'{o}}k}}, \bibinfo {author} {\bibfnamefont {R.}~\bibnamefont
  {Kondor}}, \ and\ \bibinfo {author} {\bibfnamefont {G.}~\bibnamefont
  {Cs{\'{a}}nyi}},\ }\bibfield  {title} {\enquote {\bibinfo {title} {On
  representing chemical environments},}\ }\href {\doibase
  10.1103/PhysRevB.87.184115} {\bibfield  {journal} {\bibinfo  {journal}
  {Phys.\ Rev.\ B}\ }\textbf {\bibinfo {volume} {87}},\ \bibinfo {pages}
  {184115} (\bibinfo {year} {2013})}\BibitemShut {NoStop}%
\bibitem [{\citenamefont {Deringer}\ and\ \citenamefont
  {Cs{\'{a}}nyi}(2017)}]{deri-csan17prb}%
  \BibitemOpen
  \bibfield  {author} {\bibinfo {author} {\bibfnamefont {V.~L.}\ \bibnamefont
  {Deringer}}\ and\ \bibinfo {author} {\bibfnamefont {G.}~\bibnamefont
  {Cs{\'{a}}nyi}},\ }\bibfield  {title} {\enquote {\bibinfo {title} {Machine
  learning based interatomic potential for amorphous carbon},}\ }\href
  {\doibase 10.1103/PhysRevB.95.094203} {\bibfield  {journal} {\bibinfo
  {journal} {Phys.\ Rev.\ B}\ }\textbf {\bibinfo {volume} {95}},\ \bibinfo
  {pages} {094203} (\bibinfo {year} {2017})}\BibitemShut {NoStop}%
\bibitem [{\citenamefont {Szlachta}\ \emph {et~al.}(2014)\citenamefont
  {Szlachta}, \citenamefont {Bart{\'{o}}k},\ and\ \citenamefont
  {Cs{\'{a}}nyi}}]{szla+14prb}%
  \BibitemOpen
  \bibfield  {author} {\bibinfo {author} {\bibfnamefont {W.~J.}\ \bibnamefont
  {Szlachta}}, \bibinfo {author} {\bibfnamefont {A.~P.}\ \bibnamefont
  {Bart{\'{o}}k}}, \ and\ \bibinfo {author} {\bibfnamefont {G.}~\bibnamefont
  {Cs{\'{a}}nyi}},\ }\bibfield  {title} {\enquote {\bibinfo {title} {Accuracy
  and transferability of {G}aussian approximation potential models for
  tungsten},}\ }\href {\doibase 10.1103/PhysRevB.90.104108} {\bibfield
  {journal} {\bibinfo  {journal} {Phys.\ Rev.\ B}\ }\textbf {\bibinfo {volume}
  {90}},\ \bibinfo {pages} {104108} (\bibinfo {year} {2014})}\BibitemShut
  {NoStop}%
\bibitem [{\citenamefont {Bart{\'{o}}k}\ \emph {et~al.}(2017)\citenamefont
  {Bart{\'{o}}k}, \citenamefont {De}, \citenamefont {Poelking}, \citenamefont
  {Bernstein}, \citenamefont {Kermode}, \citenamefont {Cs{\'{a}}nyi},\ and\
  \citenamefont {Ceriotti}}]{bart+17sa}%
  \BibitemOpen
  \bibfield  {author} {\bibinfo {author} {\bibfnamefont {A.~P.}\ \bibnamefont
  {Bart{\'{o}}k}}, \bibinfo {author} {\bibfnamefont {S.}~\bibnamefont {De}},
  \bibinfo {author} {\bibfnamefont {C.}~\bibnamefont {Poelking}}, \bibinfo
  {author} {\bibfnamefont {N.}~\bibnamefont {Bernstein}}, \bibinfo {author}
  {\bibfnamefont {J.}~\bibnamefont {Kermode}}, \bibinfo {author} {\bibfnamefont
  {G.}~\bibnamefont {Cs{\'{a}}nyi}}, \ and\ \bibinfo {author} {\bibfnamefont
  {M.}~\bibnamefont {Ceriotti}},\ }\bibfield  {title} {\enquote {\bibinfo
  {title} {Machine learning unifies the modelling of materials and
  molecules},}\ }\href {\doibase 10.1126/sciadv.1701816} {\bibfield  {journal}
  {\bibinfo  {journal} {Sci.\ Adv.}\ }\textbf {\bibinfo {volume} {3}},\
  \bibinfo {pages} {e1701816} (\bibinfo {year} {2017})}\BibitemShut {NoStop}%
\bibitem [{\citenamefont {Musil}\ \emph {et~al.}(2018)\citenamefont {Musil},
  \citenamefont {De}, \citenamefont {Yang}, \citenamefont {Campbell},
  \citenamefont {Day},\ and\ \citenamefont {Ceriotti}}]{musi+18cs}%
  \BibitemOpen
  \bibfield  {author} {\bibinfo {author} {\bibfnamefont {F.}~\bibnamefont
  {Musil}}, \bibinfo {author} {\bibfnamefont {S.}~\bibnamefont {De}}, \bibinfo
  {author} {\bibfnamefont {J.}~\bibnamefont {Yang}}, \bibinfo {author}
  {\bibfnamefont {J.~E.}\ \bibnamefont {Campbell}}, \bibinfo {author}
  {\bibfnamefont {G.~M.}\ \bibnamefont {Day}}, \ and\ \bibinfo {author}
  {\bibfnamefont {M.}~\bibnamefont {Ceriotti}},\ }\bibfield  {title} {\enquote
  {\bibinfo {title} {Machine learning for the structure--energy--property
  landscapes of molecular crystals},}\ }\href {\doibase 10.1039/c7sc04665k}
  {\bibfield  {journal} {\bibinfo  {journal} {Chem.\ Sci.}\ }\textbf {\bibinfo
  {volume} {9}},\ \bibinfo {pages} {1289--1300} (\bibinfo {year}
  {2018})}\BibitemShut {NoStop}%
\bibitem [{\citenamefont {{Himanen}}\ \emph {et~al.}(2019)\citenamefont
  {{Himanen}}, \citenamefont {{J{\"a}ger}}, \citenamefont {{Morooka}},
  \citenamefont {{Federici Canova}}, \citenamefont {{Ranawat}}, \citenamefont
  {{Gao}}, \citenamefont {{Rinke}},\ and\ \citenamefont {{Foster}}}]{dscribe}%
  \BibitemOpen
  \bibfield  {author} {\bibinfo {author} {\bibfnamefont {L.}~\bibnamefont
  {{Himanen}}}, \bibinfo {author} {\bibfnamefont {M.~O.~J.}\ \bibnamefont
  {{J{\"a}ger}}}, \bibinfo {author} {\bibfnamefont {E.~V.}\ \bibnamefont
  {{Morooka}}}, \bibinfo {author} {\bibfnamefont {F.}~\bibnamefont {{Federici
  Canova}}}, \bibinfo {author} {\bibfnamefont {Y.~S.}\ \bibnamefont
  {{Ranawat}}}, \bibinfo {author} {\bibfnamefont {D.~Z.}\ \bibnamefont
  {{Gao}}}, \bibinfo {author} {\bibfnamefont {P.}~\bibnamefont {{Rinke}}}, \
  and\ \bibinfo {author} {\bibfnamefont {A.~S.}\ \bibnamefont {{Foster}}},\
  }\bibfield  {title} {\enquote {\bibinfo {title} {{DScribe}: {L}ibrary of
  descriptors for machine learning in materials science},}\ }\href@noop {}
  {\bibfield  {journal} {\bibinfo  {journal} {arXiv preprint}\ } (\bibinfo
  {year} {2019})},\ \Eprint {http://arxiv.org/abs/1904.08875} {1904.08875}
  \BibitemShut {NoStop}%
\bibitem [{\citenamefont {Sch{\"o}lkopf}\ \emph {et~al.}(1998)\citenamefont
  {Sch{\"o}lkopf}, \citenamefont {Smola},\ and\ \citenamefont
  {M{\"u}ller}}]{scholkopf1998nonlinear}%
  \BibitemOpen
  \bibfield  {author} {\bibinfo {author} {\bibfnamefont {B.}~\bibnamefont
  {Sch{\"o}lkopf}}, \bibinfo {author} {\bibfnamefont {A.}~\bibnamefont
  {Smola}}, \ and\ \bibinfo {author} {\bibfnamefont {K.-R.}\ \bibnamefont
  {M{\"u}ller}},\ }\bibfield  {title} {\enquote {\bibinfo {title} {Nonlinear
  component analysis as a kernel eigenvalue problem},}\ }\href {\doibase
  10.1162/089976698300017467} {\bibfield  {journal} {\bibinfo  {journal}
  {Neural Comput.}\ }\textbf {\bibinfo {volume} {10}},\ \bibinfo {pages}
  {1299--1319} (\bibinfo {year} {1998})}\BibitemShut {NoStop}%
\bibitem [{\citenamefont {Ester}\ \emph {et~al.}(1996)\citenamefont {Ester},
  \citenamefont {Kriegel}, \citenamefont {Sander},\ and\ \citenamefont
  {Xu}}]{ester1996density}%
  \BibitemOpen
  \bibfield  {author} {\bibinfo {author} {\bibfnamefont {M.}~\bibnamefont
  {Ester}}, \bibinfo {author} {\bibfnamefont {H.-P.}\ \bibnamefont {Kriegel}},
  \bibinfo {author} {\bibfnamefont {J.}~\bibnamefont {Sander}}, \ and\ \bibinfo
  {author} {\bibfnamefont {X.}~\bibnamefont {Xu}},\ }\bibfield  {title}
  {\enquote {\bibinfo {title} {A density-based algorithm for discovering
  clusters in large spatial databases with noise},}\ }in\ \href
  {https://www.aaai.org/Papers/KDD/1996/KDD96-037.pdf} {\emph {\bibinfo
  {booktitle} {Proceedings of the Second International Conference on Knowledge
  Discovery and Data Mining}}}\ (\bibinfo  {publisher} {AAAI Press},\ \bibinfo
  {year} {1996})\ pp.\ \bibinfo {pages} {226--231}\BibitemShut {NoStop}%
\bibitem [{\citenamefont {Lyle}\ \emph {et~al.}(2015)\citenamefont {Lyle},
  \citenamefont {Pickard},\ and\ \citenamefont {Needs}}]{lyle2015prediction}%
  \BibitemOpen
  \bibfield  {author} {\bibinfo {author} {\bibfnamefont {M.~J.}\ \bibnamefont
  {Lyle}}, \bibinfo {author} {\bibfnamefont {C.~J.}\ \bibnamefont {Pickard}}, \
  and\ \bibinfo {author} {\bibfnamefont {R.~J.}\ \bibnamefont {Needs}},\
  }\bibfield  {title} {\enquote {\bibinfo {title} {Prediction of 10-fold
  coordinated \ce{TiO2} and \ce{SiO2} structures at multimegabar pressures},}\
  }\href {\doibase 10.1073/pnas.1500604112} {\bibfield  {journal} {\bibinfo
  {journal} {Proc.\ Natl Acad.\ Sci.\ U.~S.~A.}\ }\textbf {\bibinfo {volume}
  {112}},\ \bibinfo {pages} {6898--6901} (\bibinfo {year} {2015})}\BibitemShut
  {NoStop}%
\bibitem [{\citenamefont {Collins}\ and\ \citenamefont
  {Smith}(1996)}]{Collins1996b}%
  \BibitemOpen
  \bibfield  {author} {\bibinfo {author} {\bibfnamefont {D.~R.}\ \bibnamefont
  {Collins}}\ and\ \bibinfo {author} {\bibfnamefont {W.}~\bibnamefont
  {Smith}},\ }\href
  {https://epubs.stfc.ac.uk/manifestation/1174/DL-TR-96-001.pdf} {\emph
  {\bibinfo {title} {Evaluation of \ce{TiO2} force fields}}},\ \bibinfo {type}
  {Tech. Rep.}\ \bibinfo {number} {DL-TR-96-001}\ (\bibinfo  {institution}
  {Council for the Central Laboratory of the Research Councils},\ \bibinfo
  {year} {1996})\BibitemShut {NoStop}%
\bibitem [{\citenamefont {Swamy}\ \emph {et~al.}(2001)\citenamefont {Swamy},
  \citenamefont {Gale},\ and\ \citenamefont {Dubrovinsky}}]{Swamy2001}%
  \BibitemOpen
  \bibfield  {author} {\bibinfo {author} {\bibfnamefont {V.}~\bibnamefont
  {Swamy}}, \bibinfo {author} {\bibfnamefont {J.~D.}\ \bibnamefont {Gale}}, \
  and\ \bibinfo {author} {\bibfnamefont {L.~S.}\ \bibnamefont {Dubrovinsky}},\
  }\bibfield  {title} {\enquote {\bibinfo {title} {Atomistic simulation of the
  crystal structures and bulk moduli of {{\ce{TiO2}}} polymorphs},}\ }\href
  {\doibase 10.1016/S0022-3697(00)00246-8} {\bibfield  {journal} {\bibinfo
  {journal} {J.\ Phys.\ Chem.\ Solids}\ }\textbf {\bibinfo {volume} {62}},\
  \bibinfo {pages} {887--895} (\bibinfo {year} {2001})}\BibitemShut {NoStop}%
\bibitem [{\citenamefont {Trail}\ \emph {et~al.}(2017)\citenamefont {Trail},
  \citenamefont {Monserrat}, \citenamefont {R{\'\i}os}, \citenamefont
  {Maezono},\ and\ \citenamefont {Needs}}]{trail2017quantum}%
  \BibitemOpen
  \bibfield  {author} {\bibinfo {author} {\bibfnamefont {J.}~\bibnamefont
  {Trail}}, \bibinfo {author} {\bibfnamefont {B.}~\bibnamefont {Monserrat}},
  \bibinfo {author} {\bibfnamefont {P.~L.}\ \bibnamefont {R{\'\i}os}}, \bibinfo
  {author} {\bibfnamefont {R.}~\bibnamefont {Maezono}}, \ and\ \bibinfo
  {author} {\bibfnamefont {R.~J.}\ \bibnamefont {Needs}},\ }\bibfield  {title}
  {\enquote {\bibinfo {title} {Quantum monte carlo study of the energetics of
  the rutile, anatase, brookite, and columbite \ce{TiO2} polymorphs},}\ }\href
  {\doibase 10.1103/PhysRevB.95.121108} {\bibfield  {journal} {\bibinfo
  {journal} {Phys.\ Rev.\ B}\ }\textbf {\bibinfo {volume} {95}},\ \bibinfo
  {pages} {121108} (\bibinfo {year} {2017})}\BibitemShut {NoStop}%
\bibitem [{\citenamefont {Ma}\ \emph {et~al.}(2009)\citenamefont {Ma},
  \citenamefont {Liang}, \citenamefont {Miao}, \citenamefont {Bie},
  \citenamefont {Zhang}, \citenamefont {Xu},\ and\ \citenamefont
  {Jiang}}]{Ma2009}%
  \BibitemOpen
  \bibfield  {author} {\bibinfo {author} {\bibfnamefont {X.~G.}\ \bibnamefont
  {Ma}}, \bibinfo {author} {\bibfnamefont {P.}~\bibnamefont {Liang}}, \bibinfo
  {author} {\bibfnamefont {L.}~\bibnamefont {Miao}}, \bibinfo {author}
  {\bibfnamefont {S.~W.}\ \bibnamefont {Bie}}, \bibinfo {author} {\bibfnamefont
  {C.~K.}\ \bibnamefont {Zhang}}, \bibinfo {author} {\bibfnamefont
  {L.}~\bibnamefont {Xu}}, \ and\ \bibinfo {author} {\bibfnamefont {J.~J.}\
  \bibnamefont {Jiang}},\ }\bibfield  {title} {\enquote {\bibinfo {title}
  {Pressure-induced phase transition and elastic properties of {{\ce{TiO2}}}
  polymorphs},}\ }\href {\doibase 10.1002/pssb.200945111} {\bibfield  {journal}
  {\bibinfo  {journal} {Phys.\ Status Solidi B}\ }\textbf {\bibinfo {volume}
  {246}},\ \bibinfo {pages} {2132--2139} (\bibinfo {year} {2009})}\BibitemShut
  {NoStop}%
\bibitem [{\citenamefont {Mei}\ \emph {et~al.}(2014)\citenamefont {Mei},
  \citenamefont {Wang}, \citenamefont {Shang},\ and\ \citenamefont
  {Liu}}]{Mei2014}%
  \BibitemOpen
  \bibfield  {author} {\bibinfo {author} {\bibfnamefont {Z.-G.}\ \bibnamefont
  {Mei}}, \bibinfo {author} {\bibfnamefont {Y.}~\bibnamefont {Wang}}, \bibinfo
  {author} {\bibfnamefont {S.}~\bibnamefont {Shang}}, \ and\ \bibinfo {author}
  {\bibfnamefont {Z.-K.}\ \bibnamefont {Liu}},\ }\bibfield  {title} {\enquote
  {\bibinfo {title} {First-principles study of the mechanical properties and
  phase stability of \ce{TiO2}},}\ }\href {\doibase
  10.1016/j.commatsci.2013.11.020} {\bibfield  {journal} {\bibinfo  {journal}
  {Comput.\ Mater.\ Sci.}\ }\textbf {\bibinfo {volume} {83}},\ \bibinfo {pages}
  {114--119} (\bibinfo {year} {2014})}\BibitemShut {NoStop}%
\bibitem [{\citenamefont {Zhu}\ and\ \citenamefont {Gao}(2014)}]{Zhu2014}%
  \BibitemOpen
  \bibfield  {author} {\bibinfo {author} {\bibfnamefont {T.}~\bibnamefont
  {Zhu}}\ and\ \bibinfo {author} {\bibfnamefont {S.-P.}\ \bibnamefont {Gao}},\
  }\bibfield  {title} {\enquote {\bibinfo {title} {The stability, electronic
  structure, and optical property of \ce{TiO2} polymorphs},}\ }\href {\doibase
  10.1021/jp412462m} {\bibfield  {journal} {\bibinfo  {journal} {J.\ Phys.\
  Chem.~C}\ }\textbf {\bibinfo {volume} {118}},\ \bibinfo {pages}
  {11385--11396} (\bibinfo {year} {2014})}\BibitemShut {NoStop}%
\bibitem [{\citenamefont {Xu}\ \emph {et~al.}(2005)\citenamefont {Xu},
  \citenamefont {Blawzdziewicz},\ and\ \citenamefont {O'Hern}}]{xu2005random}%
  \BibitemOpen
  \bibfield  {author} {\bibinfo {author} {\bibfnamefont {N.}~\bibnamefont
  {Xu}}, \bibinfo {author} {\bibfnamefont {J.}~\bibnamefont {Blawzdziewicz}}, \
  and\ \bibinfo {author} {\bibfnamefont {C.~S.}\ \bibnamefont {O'Hern}},\
  }\bibfield  {title} {\enquote {\bibinfo {title} {Random close packing
  revisited: {W}ays to pack frictionless disks},}\ }\href {\doibase
  10.1103/PhysRevE.71.061306} {\bibfield  {journal} {\bibinfo  {journal}
  {Phys.\ Rev.\ E}\ }\textbf {\bibinfo {volume} {71}},\ \bibinfo {pages}
  {061306} (\bibinfo {year} {2005})}\BibitemShut {NoStop}%
\bibitem [{\citenamefont {Asenjo}\ \emph {et~al.}(2014)\citenamefont {Asenjo},
  \citenamefont {Paillusson},\ and\ \citenamefont
  {Frenkel}}]{asenjo2014numerical}%
  \BibitemOpen
  \bibfield  {author} {\bibinfo {author} {\bibfnamefont {D.}~\bibnamefont
  {Asenjo}}, \bibinfo {author} {\bibfnamefont {F.}~\bibnamefont {Paillusson}},
  \ and\ \bibinfo {author} {\bibfnamefont {D.}~\bibnamefont {Frenkel}},\
  }\bibfield  {title} {\enquote {\bibinfo {title} {Numerical calculation of
  granular entropy},}\ }\href {\doibase 10.1103/PhysRevLett.112.098002}
  {\bibfield  {journal} {\bibinfo  {journal} {Phys.\ Rev.\ Lett.}\ }\textbf
  {\bibinfo {volume} {112}},\ \bibinfo {pages} {098002} (\bibinfo {year}
  {2014})}\BibitemShut {NoStop}%
\bibitem [{\citenamefont {Martiniani}\ \emph {et~al.}(2017)\citenamefont
  {Martiniani}, \citenamefont {Schrenk}, \citenamefont {Ramola}, \citenamefont
  {Chakraborty},\ and\ \citenamefont {Frenkel}}]{martiniani2017numerical}%
  \BibitemOpen
  \bibfield  {author} {\bibinfo {author} {\bibfnamefont {S.}~\bibnamefont
  {Martiniani}}, \bibinfo {author} {\bibfnamefont {K.~J.}\ \bibnamefont
  {Schrenk}}, \bibinfo {author} {\bibfnamefont {K.}~\bibnamefont {Ramola}},
  \bibinfo {author} {\bibfnamefont {B.}~\bibnamefont {Chakraborty}}, \ and\
  \bibinfo {author} {\bibfnamefont {D.}~\bibnamefont {Frenkel}},\ }\bibfield
  {title} {\enquote {\bibinfo {title} {Numerical test of the {E}dwards
  conjecture shows that all packings are equally probable at jamming},}\ }\href
  {\doibase 10.1038/nphys4168} {\bibfield  {journal} {\bibinfo  {journal}
  {Nat.\ Phys.}\ }\textbf {\bibinfo {volume} {13}},\ \bibinfo {pages}
  {848--851} (\bibinfo {year} {2017})}\BibitemShut {NoStop}%
\bibitem [{\citenamefont {Martiniani}\ \emph {et~al.}(2016)\citenamefont
  {Martiniani}, \citenamefont {Schrenk}, \citenamefont {Stevenson},
  \citenamefont {Wales},\ and\ \citenamefont {Frenkel}}]{Martiniani2016}%
  \BibitemOpen
  \bibfield  {author} {\bibinfo {author} {\bibfnamefont {S.}~\bibnamefont
  {Martiniani}}, \bibinfo {author} {\bibfnamefont {K.~J.}\ \bibnamefont
  {Schrenk}}, \bibinfo {author} {\bibfnamefont {J.~D.}\ \bibnamefont
  {Stevenson}}, \bibinfo {author} {\bibfnamefont {D.~J.}\ \bibnamefont
  {Wales}}, \ and\ \bibinfo {author} {\bibfnamefont {D.}~\bibnamefont
  {Frenkel}},\ }\bibfield  {title} {\enquote {\bibinfo {title} {Turning
  intractable counting into sampling: {C}omputing the configurational entropy
  of three-dimensional jammed packings},}\ }\href {\doibase
  10.1103/PhysRevE.93.012906} {\bibfield  {journal} {\bibinfo  {journal}
  {Phys.\ Rev.~E}\ }\textbf {\bibinfo {volume} {93}},\ \bibinfo {pages}
  {012906} (\bibinfo {year} {2016})}\BibitemShut {NoStop}%
\bibitem [{\citenamefont {Daza}\ \emph {et~al.}(2016)\citenamefont {Daza},
  \citenamefont {Wagemakers}, \citenamefont {Georgeot}, \citenamefont
  {Gu{\'{e}}ry-Odelin},\ and\ \citenamefont {Sanju{\'{a}}n}}]{Daza2016}%
  \BibitemOpen
  \bibfield  {author} {\bibinfo {author} {\bibfnamefont {A.}~\bibnamefont
  {Daza}}, \bibinfo {author} {\bibfnamefont {A.}~\bibnamefont {Wagemakers}},
  \bibinfo {author} {\bibfnamefont {B.}~\bibnamefont {Georgeot}}, \bibinfo
  {author} {\bibfnamefont {D.}~\bibnamefont {Gu{\'{e}}ry-Odelin}}, \ and\
  \bibinfo {author} {\bibfnamefont {M.~A.~F.}\ \bibnamefont {Sanju{\'{a}}n}},\
  }\bibfield  {title} {\enquote {\bibinfo {title} {Basin entropy: {A} new tool
  to analyze uncertainty in dynamical systems},}\ }\href {\doibase
  10.1038/srep31416} {\bibfield  {journal} {\bibinfo  {journal} {Sci.\ Rep.}\
  }\textbf {\bibinfo {volume} {6}},\ \bibinfo {pages} {31416} (\bibinfo {year}
  {2016})}\BibitemShut {NoStop}%
\bibitem [{\citenamefont {Stevanovi{\'{c}}}(2016)}]{Stevanovi2016}%
  \BibitemOpen
  \bibfield  {author} {\bibinfo {author} {\bibfnamefont {V.}~\bibnamefont
  {Stevanovi{\'{c}}}},\ }\bibfield  {title} {\enquote {\bibinfo {title}
  {Sampling polymorphs of ionic solids using random superlattices},}\ }\href
  {\doibase 10.1103/physrevlett.116.075503} {\bibfield  {journal} {\bibinfo
  {journal} {Phys.\ Rev.\ Lett.}\ }\textbf {\bibinfo {volume} {116}},\ \bibinfo
  {pages} {075503} (\bibinfo {year} {2016})}\BibitemShut {NoStop}%
\bibitem [{\citenamefont {Musil}\ \emph {et~al.}(2019)\citenamefont {Musil},
  \citenamefont {Willatt}, \citenamefont {Langovoy},\ and\ \citenamefont
  {Ceriotti}}]{musil2019fast}%
  \BibitemOpen
  \bibfield  {author} {\bibinfo {author} {\bibfnamefont {F.}~\bibnamefont
  {Musil}}, \bibinfo {author} {\bibfnamefont {M.~J.}\ \bibnamefont {Willatt}},
  \bibinfo {author} {\bibfnamefont {M.~A.}\ \bibnamefont {Langovoy}}, \ and\
  \bibinfo {author} {\bibfnamefont {M.}~\bibnamefont {Ceriotti}},\ }\bibfield
  {title} {\enquote {\bibinfo {title} {Fast and accurate uncertainty estimation
  in chemical machine learning},}\ }\href {\doibase 10.1021/acs.jctc.8b00959}
  {\bibfield  {journal} {\bibinfo  {journal} {J.\ Chem.\ Theory Comput.}\
  }\textbf {\bibinfo {volume} {15}},\ \bibinfo {pages} {906--915} (\bibinfo
  {year} {2019})}\BibitemShut {NoStop}%
\bibitem [{\citenamefont {Faber}\ \emph {et~al.}(2017)\citenamefont {Faber},
  \citenamefont {Hutchison}, \citenamefont {Huang}, \citenamefont {Gilmer},
  \citenamefont {Schoenholz}, \citenamefont {Dahl}, \citenamefont {Vinyals},
  \citenamefont {Kearnes}, \citenamefont {Riley},\ and\ \citenamefont
  {Von~Lilienfeld}}]{faber2017prediction}%
  \BibitemOpen
  \bibfield  {author} {\bibinfo {author} {\bibfnamefont {F.~A.}\ \bibnamefont
  {Faber}}, \bibinfo {author} {\bibfnamefont {L.}~\bibnamefont {Hutchison}},
  \bibinfo {author} {\bibfnamefont {B.}~\bibnamefont {Huang}}, \bibinfo
  {author} {\bibfnamefont {J.}~\bibnamefont {Gilmer}}, \bibinfo {author}
  {\bibfnamefont {S.~S.}\ \bibnamefont {Schoenholz}}, \bibinfo {author}
  {\bibfnamefont {G.~E.}\ \bibnamefont {Dahl}}, \bibinfo {author}
  {\bibfnamefont {O.}~\bibnamefont {Vinyals}}, \bibinfo {author} {\bibfnamefont
  {S.}~\bibnamefont {Kearnes}}, \bibinfo {author} {\bibfnamefont {P.~F.}\
  \bibnamefont {Riley}}, \ and\ \bibinfo {author} {\bibfnamefont {O.~A.}\
  \bibnamefont {Von~Lilienfeld}},\ }\bibfield  {title} {\enquote {\bibinfo
  {title} {Prediction errors of molecular machine learning models lower than
  hybrid {DFT} error},}\ }\href {\doibase 10.1021/acs.jctc.7b00577} {\bibfield
  {journal} {\bibinfo  {journal} {J.\ Chem.\ Theory Comput.}\ }\textbf
  {\bibinfo {volume} {13}},\ \bibinfo {pages} {5255--5264} (\bibinfo {year}
  {2017})}\BibitemShut {NoStop}%
\bibitem [{\citenamefont {Chen}\ \emph {et~al.}(2019)\citenamefont {Chen},
  \citenamefont {Ye}, \citenamefont {Zuo}, \citenamefont {Zheng},\ and\
  \citenamefont {Ong}}]{Chen2019}%
  \BibitemOpen
  \bibfield  {author} {\bibinfo {author} {\bibfnamefont {C.}~\bibnamefont
  {Chen}}, \bibinfo {author} {\bibfnamefont {W.}~\bibnamefont {Ye}}, \bibinfo
  {author} {\bibfnamefont {Y.}~\bibnamefont {Zuo}}, \bibinfo {author}
  {\bibfnamefont {C.}~\bibnamefont {Zheng}}, \ and\ \bibinfo {author}
  {\bibfnamefont {S.~P.}\ \bibnamefont {Ong}},\ }\bibfield  {title} {\enquote
  {\bibinfo {title} {Graph networks as a universal machine learning framework
  for molecules and crystals},}\ }\href {\doibase
  10.1021/acs.chemmater.9b01294} {\bibfield  {journal} {\bibinfo  {journal}
  {Chem. Mater.}\ }\textbf {\bibinfo {volume} {31}},\ \bibinfo {pages}
  {3564--3572} (\bibinfo {year} {2019})}\BibitemShut {NoStop}%
\bibitem [{\citenamefont {Bartel}\ \emph {et~al.}(2018)\citenamefont {Bartel},
  \citenamefont {Millican}, \citenamefont {Deml}, \citenamefont {Rumptz},
  \citenamefont {Tumas}, \citenamefont {Weimer}, \citenamefont {Lany},
  \citenamefont {Stevanovi{\'{c}}}, \citenamefont {Musgrave},\ and\
  \citenamefont {Holder}}]{Bartel2018}%
  \BibitemOpen
  \bibfield  {author} {\bibinfo {author} {\bibfnamefont {C.~J.}\ \bibnamefont
  {Bartel}}, \bibinfo {author} {\bibfnamefont {S.~L.}\ \bibnamefont
  {Millican}}, \bibinfo {author} {\bibfnamefont {A.~M.}\ \bibnamefont {Deml}},
  \bibinfo {author} {\bibfnamefont {J.~R.}\ \bibnamefont {Rumptz}}, \bibinfo
  {author} {\bibfnamefont {W.}~\bibnamefont {Tumas}}, \bibinfo {author}
  {\bibfnamefont {A.~W.}\ \bibnamefont {Weimer}}, \bibinfo {author}
  {\bibfnamefont {S.}~\bibnamefont {Lany}}, \bibinfo {author} {\bibfnamefont
  {V.}~\bibnamefont {Stevanovi{\'{c}}}}, \bibinfo {author} {\bibfnamefont
  {C.~B.}\ \bibnamefont {Musgrave}}, \ and\ \bibinfo {author} {\bibfnamefont
  {A.~M.}\ \bibnamefont {Holder}},\ }\bibfield  {title} {\enquote {\bibinfo
  {title} {Physical descriptor for the {G}ibbs energy of inorganic crystalline
  solids and temperature-dependent materials chemistry},}\ }\href {\doibase
  10.1038/s41467-018-06682-4} {\bibfield  {journal} {\bibinfo  {journal} {Nat.\
  Commun.}\ }\textbf {\bibinfo {volume} {9}} (\bibinfo {year} {2018}),\
  10.1038/s41467-018-06682-4}\BibitemShut {NoStop}%
\bibitem [{Note1()}]{Note1}%
  \BibitemOpen
  \bibinfo {note} {{\color {blue} It should however be noted that at high
  temperatures, intrinsic defects may become important~\cite
  {Nowotny2015,Sarkar2019}. Some defects, such as Frenkel-style interstitials,
  can be accounted for with our approach, but the most important defects in
  TiO$_2${} include charge transfer and cannot be modelled with simple
  empirical potentials. Given the large typical free energies of defect
  formation of all main types even at such high temperatures~\cite
  {Nowotny2015}, it is not likely that they would significantly change the
  phase behaviour; however, an investigation of this behaviour with more
  advanced potentials would be particularly interesting.}}\BibitemShut {Stop}%
\bibitem [{\citenamefont {Lewis}\ and\ \citenamefont
  {Catlow}(1985)}]{Lewis1985}%
  \BibitemOpen
  \bibfield  {author} {\bibinfo {author} {\bibfnamefont {G.~V.}\ \bibnamefont
  {Lewis}}\ and\ \bibinfo {author} {\bibfnamefont {C.~R.~A.}\ \bibnamefont
  {Catlow}},\ }\bibfield  {title} {\enquote {\bibinfo {title} {Potential models
  for ionic oxides},}\ }\href {\doibase 10.1088/0022-3719/18/6/010} {\bibfield
  {journal} {\bibinfo  {journal} {J.\ Phys.\ C: Solid State Phys.}\ }\textbf
  {\bibinfo {volume} {18}},\ \bibinfo {pages} {1149--1161} (\bibinfo {year}
  {1985})}\BibitemShut {NoStop}%
\bibitem [{\citenamefont {Shimojo}\ \emph {et~al.}(1992)\citenamefont
  {Shimojo}, \citenamefont {Okabe}, \citenamefont {Tachibana}, \citenamefont
  {Kobayashi},\ and\ \citenamefont {Okazaki}}]{Shimojo1992}%
  \BibitemOpen
  \bibfield  {author} {\bibinfo {author} {\bibfnamefont {F.}~\bibnamefont
  {Shimojo}}, \bibinfo {author} {\bibfnamefont {T.}~\bibnamefont {Okabe}},
  \bibinfo {author} {\bibfnamefont {F.}~\bibnamefont {Tachibana}}, \bibinfo
  {author} {\bibfnamefont {M.}~\bibnamefont {Kobayashi}}, \ and\ \bibinfo
  {author} {\bibfnamefont {H.}~\bibnamefont {Okazaki}},\ }\bibfield  {title}
  {\enquote {\bibinfo {title} {Molecular dynamics studies of yttria stabilized
  zirconia. {I}. {S}tructure and oxygen diffusion},}\ }\href {\doibase
  10.1143/jpsj.61.2848} {\bibfield  {journal} {\bibinfo  {journal} {J.\ Phys.\
  Soc.\ Jpn}\ }\textbf {\bibinfo {volume} {61}},\ \bibinfo {pages} {2848--2857}
  (\bibinfo {year} {1992})}\BibitemShut {NoStop}%
\bibitem [{\citenamefont {Cheng}\ \emph {et~al.}(2019)\citenamefont {Cheng},
  \citenamefont {Mazzola},\ and\ \citenamefont {Ceriotti}}]{cheng2019evidence}%
  \BibitemOpen
  \bibfield  {author} {\bibinfo {author} {\bibfnamefont {B.}~\bibnamefont
  {Cheng}}, \bibinfo {author} {\bibfnamefont {G.}~\bibnamefont {Mazzola}}, \
  and\ \bibinfo {author} {\bibfnamefont {M.}~\bibnamefont {Ceriotti}},\
  }\bibfield  {title} {\enquote {\bibinfo {title} {Evidence for supercritical
  behavior of high-pressure liquid hydrogen},}\ }\href@noop {} {\bibfield
  {journal} {\bibinfo  {journal} {arXiv preprint}\ } (\bibinfo {year}
  {2019})},\ \Eprint {http://arxiv.org/abs/1906.03341} {1906.03341}
  \BibitemShut {NoStop}%
\bibitem [{\citenamefont {Murakami}\ \emph {et~al.}(2004)\citenamefont
  {Murakami}, \citenamefont {Hirose}, \citenamefont {Kawamura}, \citenamefont
  {Sata},\ and\ \citenamefont {Ohishi}}]{murakami2004post}%
  \BibitemOpen
  \bibfield  {author} {\bibinfo {author} {\bibfnamefont {M.}~\bibnamefont
  {Murakami}}, \bibinfo {author} {\bibfnamefont {K.}~\bibnamefont {Hirose}},
  \bibinfo {author} {\bibfnamefont {K.}~\bibnamefont {Kawamura}}, \bibinfo
  {author} {\bibfnamefont {N.}~\bibnamefont {Sata}}, \ and\ \bibinfo {author}
  {\bibfnamefont {Y.}~\bibnamefont {Ohishi}},\ }\bibfield  {title} {\enquote
  {\bibinfo {title} {Post-perovskite phase transition in \ce{MgSiO3}},}\ }\href
  {\doibase 10.1126/science.1095932} {\bibfield  {journal} {\bibinfo  {journal}
  {Science}\ }\textbf {\bibinfo {volume} {304}},\ \bibinfo {pages} {855--858}
  (\bibinfo {year} {2004})}\BibitemShut {NoStop}%
\bibitem [{\citenamefont {Beran}(2016)}]{beran2016modeling}%
  \BibitemOpen
  \bibfield  {author} {\bibinfo {author} {\bibfnamefont {G.~J.~O.}\
  \bibnamefont {Beran}},\ }\bibfield  {title} {\enquote {\bibinfo {title}
  {Modeling polymorphic molecular crystals with electronic structure theory},}\
  }\href {\doibase 10.1021/acs.chemrev.5b00648} {\bibfield  {journal} {\bibinfo
   {journal} {Chem.\ Rev.}\ }\textbf {\bibinfo {volume} {116}},\ \bibinfo
  {pages} {5567--5613} (\bibinfo {year} {2016})}\BibitemShut {NoStop}%
\bibitem [{\citenamefont {Nowotny}\ \emph {et~al.}(2015)\citenamefont
  {Nowotny}, \citenamefont {Alim}, \citenamefont {Bak}, \citenamefont {Idris},
  \citenamefont {Ionescu}, \citenamefont {Prince}, \citenamefont {Sahdan},
  \citenamefont {Sopian}, \citenamefont {Mat~Teridi},\ and\ \citenamefont
  {Sigmund}}]{Nowotny2015}%
  \BibitemOpen
  \bibfield  {author} {\bibinfo {author} {\bibfnamefont {J.}~\bibnamefont
  {Nowotny}}, \bibinfo {author} {\bibfnamefont {M.~A.}\ \bibnamefont {Alim}},
  \bibinfo {author} {\bibfnamefont {T.}~\bibnamefont {Bak}}, \bibinfo {author}
  {\bibfnamefont {M.~A.}\ \bibnamefont {Idris}}, \bibinfo {author}
  {\bibfnamefont {M.}~\bibnamefont {Ionescu}}, \bibinfo {author} {\bibfnamefont
  {K.}~\bibnamefont {Prince}}, \bibinfo {author} {\bibfnamefont {M.~Z.}\
  \bibnamefont {Sahdan}}, \bibinfo {author} {\bibfnamefont {K.}~\bibnamefont
  {Sopian}}, \bibinfo {author} {\bibfnamefont {M.~A.}\ \bibnamefont
  {Mat~Teridi}}, \ and\ \bibinfo {author} {\bibfnamefont {W.}~\bibnamefont
  {Sigmund}},\ }\bibfield  {title} {\enquote {\bibinfo {title} {Defect
  chemistry and defect engineering of {{\ce{TiO2}}}-based semiconductors for
  solar energy conversion},}\ }\href {\doibase 10.1039/c4cs00469h} {\bibfield
  {journal} {\bibinfo  {journal} {Chem.\ Soc.\ Rev.}\ }\textbf {\bibinfo
  {volume} {44}},\ \bibinfo {pages} {8424--8442} (\bibinfo {year}
  {2015})}\BibitemShut {NoStop}%
\bibitem [{\citenamefont {Sarkar}\ and\ \citenamefont
  {Khan}(2019)}]{Sarkar2019}%
  \BibitemOpen
  \bibfield  {author} {\bibinfo {author} {\bibfnamefont {A.}~\bibnamefont
  {Sarkar}}\ and\ \bibinfo {author} {\bibfnamefont {G.~G.}\ \bibnamefont
  {Khan}},\ }\bibfield  {title} {\enquote {\bibinfo {title} {The formation and
  detection techniques of oxygen vacancies in titanium oxide-based
  nanostructures},}\ }\href {\doibase 10.1039/c8nr09666j} {\bibfield  {journal}
  {\bibinfo  {journal} {Nanoscale}\ }\textbf {\bibinfo {volume} {11}},\
  \bibinfo {pages} {3414--3444} (\bibinfo {year} {2019})}\BibitemShut {NoStop}%
\end{thebibliography}
\end{document}